\documentclass[10pt]{article}
\pdfoutput=1

\usepackage{amsthm,amsfonts,amsmath}
\usepackage{stackrel,amssymb}
\usepackage{pxfonts}
\usepackage{euscript}
\usepackage{hyperref}
\usepackage{xcolor}
\usepackage{tcolorbox}

\usepackage{xypic}
\usepackage[all,2cell,arrow,matrix]{xy} \UseAllTwocells \SilentMatrices
\usepackage{graphicx}
\usepackage{verbatim}
\usepackage{leftidx}
\usepackage{tikz-cd}
\usepackage{tikz}
\usetikzlibrary{arrows,arrows.meta,decorations.markings,decorations.pathreplacing,calc}

\usepackage{pgfplots}
\usepgfplotslibrary{dateplot}

\usepackage{xspace}
\tikzset{->-/.style={decoration={markings,mark=at position #1 with {\arrow{stealth}}},postaction={decorate}},->-/.default=0.6}

\usepackage{epstopdf} 
\usepackage{enumerate}
\usepackage{bbold}

\usepackage{sectsty}
\sectionfont{\Large}

\usepackage{verbatim}
\usepackage{comment}

\usepackage{appendix}

\newcommand\void[1]       {}

\theoremstyle{definition}

\newtheorem{thm}{Theorem}[section]

\newtheorem{pthm}[thm]{Theorem$^{\mathrm{ph}}$}

\theoremstyle{definition}

\newtheorem{expl}[thm]{Example}
\newtheorem{rem}[thm]{Remark}

\numberwithin{equation}{section}
\numberwithin{thm}{section}

\newcommand\nn             {\nonumber \\}
\newcommand\be            {\begin{equation}}
\newcommand\ee            {\end{equation}}
\newcommand\bea           {\begin{eqnarray}}
\newcommand\eea         {\end{eqnarray}}
\newcommand\bnu          {\begin{enumerate}}
\newcommand\enu          {\end{enumerate}}

\allowdisplaybreaks[1]

\newcommand{\pf}{\begin{proof}}
\newcommand{\epf}{\end{proof}}

\newcommand\Cb            {\mathbb{C}}

\newcommand\Zb            {\mathbb{Z}}

\newcommand\CB           {\EuScript{B}}
\newcommand\CC           {\EuScript{C}}

\newcommand\CE          {\EuScript{E}}

\newcommand\CH         {\EuScript{H}}

\newcommand\CM          {\EuScript{M}}

\newcommand\CR         {\EuScript{R}}
\newcommand\CS         {\EuScript{S}}

\newcommand\CX         {\EuScript{X}}
\newcommand\CY         {\EuScript{Y}}

\newcommand{\FZ}			{\text{\usefont{U}{euf}{m}{n}Z}}

 \DeclareMathOperator{\Aut}{Aut}

 \DeclareMathOperator{\Id}{Id}
 \DeclareMathOperator{\id}{id}

 \DeclareMathOperator{\Fun}{Fun}

 \DeclareMathOperator{\Vect}{Vect}

 \DeclareMathOperator{\Mod}{Mod}

\newcommand{\rev}{\mathrm{rev}}

\newcommand{\one}{\mathbb{1}}

\newcommand\hilb {\mathrm{Vec}}

\newcommand\vect {\mathrm{Vec}}
\newcommand\Rep {\mathrm{Rep}}

\newcommand\Pic {\mathrm{Pic}}
\newcommand\svec {\mathrm{sVec}}

\newcommand\lwll {\textsf{\small LWLL}}  

\usepackage{color}
\definecolor{red}{rgb}{1,0,0}
\definecolor{blue}{rgb}{0,0,1}
\definecolor{green}{rgb}{0,1,0}

\newcommand\arXiv[1]{\href{http://arxiv.org/abs/#1}{arXiv:#1}}

\begin{document}

\begin{center} \LARGE
One dimensional gapped quantum phases and enriched fusion categories
\end{center}

\vspace{0.01cm}
\begin{center}
Liang Kong$^{a,b,c}$,\,  
Xiao-Gang Wen$^{d}$,
Hao Zheng$^{e,f,g}$
\\[1em]
$^a$ Shenzhen Institute for Quantum Science and Engineering, \\
Southern University of Science and Technology, Shenzhen, 518055, China 
\\[0.4em]
$^b$ International Quantum Academy, 
Shenzhen 518048, China
\\[0.4em]
$^c$ Guangdong Provincial Key Laboratory of Quantum Science and Engineering, \\
Southern University of Science and Technology, Shenzhen, 518055, China
\\[0.4em]
$^d$ Department of Physics, Massachusetts Institute of Technology, \\
Cambridge, Massachusetts 02139, USA
\\[0.4em]
$^e$ Institute for Applied Mathematics, Tsinghua University, Beijing, 100084, China
\\[0.4em]
$^f$ Beijing Institute of Mathematical Sciences and Applications, Beijing 101408, China
\\[0.4em]
$^g$ Department of Mathematics, Peking University, Beijing 100871, China
\end{center}

\bigskip
\begin{abstract}

In this work, we use Ising chain and Kitaev chain to check the validity of an earlier proposal in \arXiv{2011.02859} that enriched fusion (higher) categories provide a unified categorical description of all gapped/gapless quantum liquids, including symmetry-breaking phases, topological orders, SPT/SET orders and CFT-type gapless quantum phases. In particular, we show explicitly that, in each gapped phase realized by these two models, the spacetime observables form a fusion category enriched in a braided fusion category such that its monoidal center is trivial. We also study the categorical descriptions of the boundaries of these models. In the end, we obtain the classification of and the categorical descriptions of all 1-dimensional (spatial dimension)  gapped quantum phases with a bosonic/fermionic finite onsite symmetry.

\end{abstract}

\vspace{0.05cm}
\tableofcontents

\newpage

\section{Introduction} \label{sec:introduction}
In this work, we restudy the Ising chain and the Kitaev chain from a categorical point of view. Physics oriented readers can skip this section and start from later sections directly, and come back later for the historical development of the main idea. Throughout this work, we use $n$d to represent a spatial dimension and $n+$1D to represent a spacetime dimension, and all fusion (higher) categories are assumed to be unitary \cite{KZ20b}. 

\medskip
The study of topological orders and symmetry protected/enriched topological (SPT/SET) orders has attracted a lot of attention in recent years because it goes beyond Landau's paradigm of phases and phase transitions (see \cite{Wen19} for a review and references therein). A topological order, as a macroscopic notion that defines the universal class of quantum many body systems, can be described by observables in the long wave length limit (\lwll). These observables often form categorical structures. 
For example, a 2d anomaly-free topological order can be described by the fusion and braiding structures of its particle-like topological excitations (or anyons) up to chiral central charges. These fusion-braiding structures form a unitary modular tensor category (see \cite[Appendix\ E]{Kit06} for a review). A potentially anomalous 1d topological order can be described by a unitary (multi-)fusion category \cite{KK12}. These two categorical descriptions can be checked directly from concrete lattice models (see for example \cite{Kit03,BK98,LW05,KK12,LW14,HLPWW18,CCW17}). The categorical descriptions of higher dimensional topological orders can be found in \cite{KW14,KWZ15,LKW18,LW19,KTZho20,JF20,KZ20b}. Some of them were checked in lattice models \cite{BD19,KTZha20,BD21}. 
These categorical descriptions provide a unified approach towards the classification of all topological orders.

Ever since the introduction of the notion of a SPT/SET order \cite{GW09,CGW10a,CLW11,CGLW13}, it is natural to expect that it also has a categorical description. However, the story of developing this description is full of twists and turns. We review this development in Section\,\ref{sec:SPT/SET}, and explain the main result of a unified classfication theory developed in \cite{KLWZZ20a}. However, this classification theory is still one step away from a physically natural description of an SPT/SET order. The last missing step was made in \cite{KZ20b}. Based on the idea of topological Wick rotation \cite{KZ20a}, a notion which is reviewed in Section\,\ref{sec:TWR}, two of the authors proposed in \cite[Section\ 7]{KZ21}\cite[Section\ 5.2]{KZ20b} a unified categorical description of all gapped/gapless quantum liquids\footnote{There are non-liquid quantum phases (see for example \cite{Cha05,Haa11} and \cite{ZW15}).}, including symmetry-breaking phases, topological orders, SPT/SET orders and CFT-type gapless phases, in terms of enriched higher categories. The main goal of this work is to check the validity of this proposal through concrete 1d models: the Ising chain and the Kitaev chain.


In Section\,\ref{sec:Ising-chain} and \ref{sec:Kitaev-chain}, by carefully analyzing the Ising chain and the Kitaev chain, we prove that, in each gapped phase (an SPT order or a symmetry-breaking phase) realized in these two models, observables in spacetime form a fusion category enriched in a braided fusion category such that its monoidal center is trivial. In Section\,\ref{sec:outlooks}, we provide the classification and the categorical descriptions of all 1d gapped phases with a bosonic/fermionic finite onsite symmetry. The notion of an enriched category is briefly explained in Appendix\,\ref{sec:enriched-categories}. The hom spaces of all enriched categories appeared in this work are all computed in Appendix\,\ref{sec:enriched-categories}.  



\subsection{Towards a categorical description of SPT/SET orders}  \label{sec:SPT/SET}
In 2d, a categorical description of bosonic SPT/SET orders with a finite onsite symmetry was first introduced by Barkeshli, Bonderson, Cheng and Wang in \cite{BBCW19} based on the idea of gauging the symmetry by introducing 1d symmetry defects. Later, a new description for both bosonic and fermionic 2d SPT/SET orders was introduced in \cite{LKW17a,LKW17b} also based on the idea of gauging the symmetry but in a different way \cite{LG12}. It is useful to recall the key idea of \cite{LG12,LKW17a,LKW17b}. In a 2d SPT/SET order, local (non-topological) excitations are given by the symmetry charges. They form a symmetric fusion subcategory $\CE$ in the braided fusion category $\CS$ of all local and topological excitations. Since the symmetry charges cannot be detected via double braidings in $\CS$, this can be viewed as a sign of ``anomaly'' but somehow ``protected by the symmetry'' in a not-fully-understood way, which is clarified in this work (see Remark\, \ref{rem:fix-anomaly-1} and \ref{rem:fix-anomaly-in-time}). The idea of gauging the symmetry is to introduce additional particles to $\CS$ in a minimal way such that all old and new particles can be detected by double braidings again. In mathematical language, it amounts to finding a minimal modular extension of $\CS$ \cite{Mu00,LKW17b}.

Although the idea of gauging the symmetry works and can be generalized to higher dimensions \cite{KLWZZ20a}, it is unsatisfying because the SPT/SET orders are well-defined before the gauging. There should be an intrinsic but missing data canonically associated to $\CS$ that can characterize a 2d SPT/SET order without gauging the symmetry. This dissatisfaction motivated a new description of SPT/SET orders without gauging the symmetry \cite{KLWZZ20a}. This description is based on the idea of boundary-bulk relation \cite{KWZ15,KWZ17}. More precisely, an anomaly-free $n$d SPT/SET order should have a trivial $n+$1d bulk, i.e. the trivial $n+$1d SPT order, which has a categorical description in the minimal modular extension approach. Using the fact that the bulk is the center of the boundary \cite{KWZ15,KWZ17}, we obtain a mathematical description of an anomaly-free $n$d SPT/SET order, summarized in the following physical theorem. 

\begin{pthm}\cite{KLWZZ20a} \label{thm:classification_SET}
For $n\geq 1$, let $\CR$ be a unitary symmetric fusion $n$-category viewed as a higher symmetry. 
We call an $n$d (spatial dimension) SPT/SET order with the higher symmetry $\CR$ an $n$d SPT/SET$_{/\CR}$ order. 
\bnu 
\item An anomaly-free $n$d SET$_{/\CR}$ order is uniquely (up to invertible topological orders) characterized by
a pair $(\CS,\phi)$, where $\CS$ is a unitary fusion $n$-category equipped with an embedding $\iota_\CS: \CR \hookrightarrow \FZ_1(\CS)$ such that 
\be \label{eq:condition-faithful}
\mbox{the composed functor $(\CR \hookrightarrow \FZ_1(\CS) \to \CS)$ is faithful}
\ee
and $\phi: \FZ_1(\CR) \to \FZ_1(\CS)$ is a braided equivalence between the monoidal centers of $\CR$ and $\CS$ rendering the following diagram commutative (up to a natural isomorphism):
\be \label{diag:RRA}
\xymatrix@R=0.8em{
& \CR \ar@{^(->}[dl]_{\iota_\CR} \ar@{^(->}[dr]^{\iota_\CS} & \\
\FZ_1(\CR) \ar[rr]_\simeq^\phi & & \FZ_1(\CS).
}
\ee

\item When $\CS=\CR$, the pair $(\CR, \phi)$ describes an SPT$_{/\CR}$ order and
  $(\CR,\id_{\FZ_1(\CR)})$ describes the trivial SPT$_{/\CR}$ order. Moreover, the
  group of all SPT$_{/\CR}$ orders (with the multiplication defined by the stacking and the identity element by the trivial SPT order) is isomorphic to the group
  $\Aut^{br}(\FZ_1(\CR),\iota_\CR)$, which denotes the underlying group of the braided autoequivalences of $\FZ_1(\CR)$ preserving $\iota_\CR$, i.e. $\phi\circ\iota_\CR\simeq\iota_\CR$. 
\enu
\end{pthm}

\begin{rem} \label{rem:SET+symmetry-breaking}
It was proposed later in \cite{KZ20b} that this classification should automatically includes all gapped symmetry-breaking phases if we drop the condition (\ref{eq:condition-faithful}). 
\end{rem}

\begin{figure}
\[
\begin{array}{c}
\begin{tikzpicture}[scale=0.6]
\fill[gray!20] (-4,0) rectangle (4,4) ;
\draw[->-,very thick] (0,0)--(4,0) node[midway,below] {$\CS$} ;
\draw[->-,very thick] (-4,0)--(0,0) node[midway,below] {$\CR$} ;
\draw[->-,dashed,very thick] (0,4)--(0,0) node[at start,below right,scale=0.8] {$\CY_\phi$} ;

\draw[fill=white] (-0.1,-0.1) rectangle (0.1,0.1) ;
\node[above] at (2,2) {$\FZ_1(\CS)$} ;
\node[above] at (-2,2) {$\FZ_1(\CR)$} ;

\end{tikzpicture}
\end{array}
\]
\caption{This picture depicts the physical meaning of the classification theorem of $n$d SPT/SET orders given in Theorem\,\ref{thm:classification_SET}. There are two ways to interpret this picture. One was provided in \cite{KLWZZ20a}, where $\FZ_1(\CR)$ is regarded as the 1-dimensional higher bulk of the SPT/SET order and the vertical direction is the ($n+$1)-th spatial direction. The other one was provided in \cite{KZ20b}, where the vertical direction is the time direction and $\FZ_1(\CR)$ is viewed as the background category of an enriched $n$-category ${}^{\FZ_1(\CR)}\CR$ or ${}^{\FZ_1(\CR)}\CS$ (see Appendix\,\ref{sec:enriched-categories}), the hom spaces of which encode the spacetime observables of the SPT/SET orders. 
}
\label{fig:SET}
\end{figure}
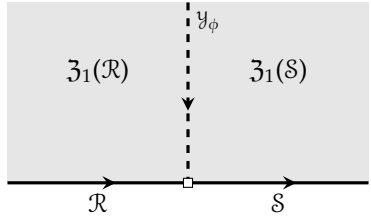

The physical meaning of Theorem\,\ref{thm:classification_SET} is illustrated in Figure\,\ref{fig:SET}. In particular, we regarded $\FZ_1(\CR)$ (resp. $\FZ_1(\CS)$) as the 1-dimensional higher bulk of the $n$d SPT (resp. SET) order, and vertical direction in Figure\,\ref{fig:SET} is the ($n+$1)-th spatial direction. The braided auto-equivalence $\phi$ in (\ref{diag:RRA}) is precisely the missing data, and can be realized physically by an invertible domain wall $\CY_\phi$ between $\FZ_1(\CR)$ and $\FZ_1(\CS)$. Note that when $\CS=\CR$, the pair $(\CR,\phi)$ describes an $n$d SPT, which is precisely realized by the $n$d invertible domain wall $\CY_\phi$.

\begin{expl} \label{expl:1d-SPT-SET}
Let $\CR=\Rep(G)$ or $\Rep(G,z)$, where $G$ is a finite group, and $\Rep(G)$ is the category of finite dimensional $G$-representations, and $\Rep(G,z)$ denotes the same category but equipped with a new symmetric braiding respecting the fermion parity $z$. All 1d anomaly-free SET$_{/\CR}$ orders are SPT$_{/\CR}$ orders. The group of 1d SPT$_{/\CR}$ orders is isomorphic to the group $\Aut^{br}(\FZ_1(\CR),\iota_\CR)$ and to the Picard group $\mathrm{Pic}(\CR)$ of $\CR$ \cite{DN13}, which was computed in \cite{Ca06}: 
\begin{align} 
\Pic(\Rep(G)) &\simeq H^2(G,U(1));  \nn
\label{eq:pic=f_0}
\Pic(\Rep(G,z)) &\simeq \left\{ \begin{array}{ll} 
H^2(G,U(1)) \times \Zb_2 & \textrm{if $G = G_b \times \langle z \rangle$}; \\
H^2(G, U(1)) & \textrm{otherwise}. 
\end{array} \right.
\end{align}
When $G=\Zb_2$, there is a unique non-trivial fermionic SPT order, which can be realized by the Kitaev chain. 
\end{expl}

\subsection{Topological Wick rotations} \label{sec:TWR}

Although Theorem\,\ref{thm:classification_SET} is successful in that it unifies all earlier classification results and is generalized to all dimensions, the classifying data given there cannot be the direct description of the observables of an SPT/SET order in \lwll. Indeed, on the one hand, the crucial data $\phi$ is associated to the categorical description of the 1-dimensional higher bulk; on the other hand, in a concrete $n$d lattice model realization of an anomaly-free $n$d SPT/SET order, its $n+$1d bulk is completely empty. Therefore, there should be a direct categorical description of the observables in an anomaly-free $n$d SPT/SET order without using its empty bulk.  

\medskip
How to find such a description? The most obvious approach is to analyze a concrete lattice model realization of an SPT/SET order, and collect all observables in \lwll~to see what mathematical structure they form. Ironically, this obvious approach has never been seriously studied. 
Perhaps, a partial reason for the delay is that, without knowing what you are looking for, it is rather difficult to walk through the labyrinth of rich ingredients in a lattice model, often misguided by old conventions and misunderstandings, to crystallize the hidden and unknown mathematical structures. In this work, we do the long-overdue homework but with a new mathematical guidance. 

The guidance came from a rather mysterious process called {\it topological Wick rotation}, which was first introduced in the study of gapless boundaries of 2d topological orders \cite[Section\ 5.2]{KZ20a}, and was generalized to all dimensions \cite[Section\ 7]{KZ21}. In a special case, it says that given an anomalous $n$d topological order, whose topological defects form a fusion $n$-category $\CS$, and its $n+$1d bulk described by the monoidal center (or the $E_1$-center) $\FZ_1(\CS)$ of $\CS$ \cite{KWZ17} (as depicted in the first picture in (\ref{pic:TWR})), one can ``rotate'' the $n+$1d bulk to the time direction to obtain an anomaly-free $n$d phase (potentially gapless) as illustrated in the second picture in (\ref{pic:TWR}). 
\be \label{pic:TWR}
\begin{array}{c}
\begin{tikzpicture}[scale=0.6]
\draw[->-,ultra thick] (0,0)--(-3,0) node[midway,above] {\scriptsize $\FZ_1(\CS)$} node[below] {\scriptsize ($n+$1)-th spatial direction};
\draw[fill=white] (-0.1,-0.1) rectangle (0.1,0.1) node[midway,above] {\scriptsize $\CS$} ;
\end{tikzpicture}
\end{array}
\quad\quad\quad\quad
\begin{array}{c}
\begin{tikzpicture}[scale=0.6]
\draw[->-,ultra thick] (0,0)--(0,3) node[midway,right] {\scriptsize $\FZ_1(\CS)$} node[left] {\scriptsize the time direction};
\draw[fill=white] (-0.1,-0.1) rectangle (0.1,0.1) node[midway,right] {\scriptsize $\CS$} ;
\end{tikzpicture}
\end{array}
\ee

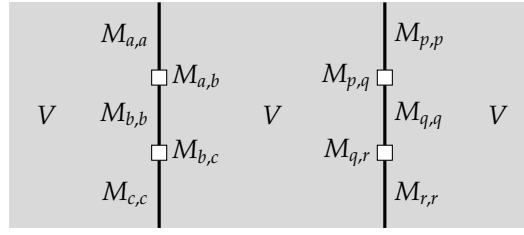
\begin{figure}
$$ 
\raisebox{-4em}{\begin{tikzpicture}
\fill[gray!30] (-3,0) rectangle (4,3) ;
\draw[very thick] (-1,3)--(-1,2.1) node[midway,left] {$M_{a,a}$};
\draw[very thick] (-1,1.9)--(-1,1.1) node[midway,left] {$M_{b,b}$};
\draw[very thick] (-1,0.9)--(-1,0) node[midway,left] {$M_{c,c}$};
\draw[fill=white] (-1.1,0.9) rectangle (-0.9,1.1) node[midway,right] {$\,M_{b,c}$} ;
\draw[fill=white] (-1.1,1.9) rectangle (-0.9,2.1) node[midway,right] {$\,M_{a,b}$} ;
\draw[very thick] (2,3)--(2,2.1) node[midway,right] {$M_{p,p}$};
\draw[very thick] (2,1.9)--(2,1.1) node[midway,right] {$M_{q,q}$};
\draw[very thick] (2,0.9)--(2,0) node[midway,right] {$M_{r,r}$};
\draw[fill=white] (1.9,0.9) rectangle (2.1,1.1) node[midway,left] {$M_{q,r}\,$} ;
\draw[fill=white] (1.9,1.9) rectangle (2.1,2.1) node[midway,left] {$M_{p,q}\,$} ;
\node at (-2.5,1.5) {$V$} ;
\node at (0.5,1.5) {$V$} ;
\node at (3.5,1.5) {$V$} ;
\end{tikzpicture}}
$$
\caption{This picture depicts all observables in \lwll~in a 1+1D CFT. In particular, $V$ denotes the local quantum symmetry, and $M_{a,b}$ is a space of defects fields, and $M_{a,a}$ defines a topological defect line (TDL). All $M_{a,b}$, together with the labels $a,b,c,\cdots \in\CS$, form an enriched category ${}^{\FZ_1(\CS)}\CS$ with $\hom_{{}^{\FZ_1(\CS)}\CS}(a,b)=M_{a,b}$. 
}
\label{fig:observables}
\end{figure}

After the topological Wick rotation, the pair $(\FZ_1(\CS),\CS)$ represents a fusion $n$-category ${}^{\FZ_1(\CS)}\CS$ enriched in $\FZ_1(\CS)$ (or a $\FZ_1(\CS)$-enriched fusion $n$-category). The enriched fusion $n$-category ${}^{\FZ_1(\CS)}\CS$ is called the {\it topological skeleton}\footnote{The topological skeleton can also be defined by $\CS$ as in \cite{KZ20b} because ${}^{\FZ_1(\CS)}\CS$ does not contain more information than $\CS$.} of the anomaly-free $n$d phase. It turns out that the topological skeleton ${}^{\FZ_1(\CS)}\CS$ does not contain all the information of the anomaly-free $n$d phase. The physical meaning of this topological skeleton is better explained together with the missing information. In a 1+1D rational CFT, the missing information is the so-called local quantum symmetry $V$, which is either a chiral symmetry (defined by a vertex operator algebra (VOA)) or a non-chiral symmetry (defined by a full field algebra \cite{HK07}), together with a braided functor $\phi: \Rep(V) \to \FZ_1(\CS)$, where $\Rep(V)$ denotes the category of $V$-representations.
In other words, the triple $(V,\phi,{}^{\FZ_1(\CS)}\CS)$ gives a complete information of a 1+1D rational CFT. The braided equivalence $\phi$ endows the abstract enriched category ${}^{\FZ_1(\CS)}\CS$ with a precise physical meaning. In particular, the objects in ${}^{\FZ_1(\CS)}\CS$ are objects in $\CS$, and are the labels of topological defect lines (TDL) admitted by the local quantum symmetry. We illustrate these TDL's and 0D defects among them in Figure\, \ref{fig:observables}. For $a,b\in\CS$, the hom space $\hom_{{}^{\FZ_1(\CS)}\CS}(a,b)$ consists of (chiral or non-chiral) fields operators\footnote{They are also called boundary-condition changing operators in CFT.} that respect the local quantum symmetry (see \cite[Section\ 3.4]{KZ20a} for more details). These cover all observables in a 1+1D CFT. 

What we mean by respecting the local quantum symmetry is that the space $\hom_{{}^{\FZ_1(\CS)}\CS}(a,b)$ of field operators is a $V$-representation, and the operator product expansion (OPE) among these operators are defined by chiral vertex operators \cite{MS89}, or more precisely, by the intertwining operators of $V$ \cite{FHL93}. By the representation theory of VOA's \cite{HL95}, it means that $\hom_{{}^{\FZ_1(\CS)}\CS}(a,b)$ can be viewed as an object in $\Rep(V)$ and all the composition maps among these hom spaces are morphisms in $\Rep(V)$. This is just another way to say that the category is enriched in $\Rep(V)$. See \cite{KZ20a,KZ21} for more details. 

\begin{rem} \label{rem:fix-anomaly-1}
In the first picture in (\ref{pic:TWR}), before the topological Wick rotation, $\CS$ is anomalous as an $n$d topological order, and the anomaly is fixed by its $n+$1d (spatial dimension) bulk. After the topological Wick rotation, the anomaly is fixed by the operators in $n+$1D spacetime. 
\end{rem}

In \cite[Section\,7]{KZ21}, two of the authors proposed to generalize above picture (including the topological Wick rotation) to higher dimensions to give a unified theory for gapped/gapless phases without knowing how to include SPT/SET orders. 
Inspired by the classification of SPT/SET orders in \cite{KLWZZ20a} and the observation that an onsite symmetry should be a special case of local quantum symmetries, two of the authors proposed in \cite[Section\ 5.2]{KZ20b} to apply the topological Wick rotation to all boundary-bulk configurations (e.g. Figure\,\ref{fig:SET}) to obtain a new description of SPT/SET orders in terms of enriched higher categories. This leads to a grand unification of all gapped/gapless quantum liquids with/without onsite symmetries (including symmetry-breaking phases). 
In particular, the results in Theorem\,\ref{thm:classification_SET} can be reinterpreted by ``rotating'' the $n+$1d bulk in Figure\,\ref{fig:SET} to the time direction and reinterpreting the pair $(\CS, \phi)$ as an $\FZ_1(\CS)$-enriched fusion $n$-category ${}^{\FZ_1(\CS)}\CS$ determined by the braided equivalence $\phi: \FZ_1(\CR) \to \FZ_1(\CS)$. In this process, the $n+$1d bulk excitations in $\FZ_1(\CS)$ before the rotation are replaced by symmetric (non-local) operators in the $n+$1D spacetime after the rotation.

\begin{expl} \label{expl:nd-SPT}
For an $n$d SPT order with a finite onsite symmetry $G$, topological excitations in $\CS$ consist of all the symmetry charges and their condensation descendants. They form a symmetric fusion $n$-category $n\Rep(G)$ in the bosonic case or $n\Rep(G,z)$ in the fermionic case \cite{KLWZZ20a}, where $z\in G$ is the fermion parity. Therefore, an $n$d SPT order with the bosonic onsite symmetry $G$ should be categorically described by an enriched fusion $n$-category ${}^{\FZ_1(n\Rep(G))}\CS$ for $\CS=n\Rep(G)$, which is defined by a braided equivalence $\phi: \FZ_1(n\Rep(G)) \to \FZ_1(\CS)$ preserving $n\Rep(G) \hookrightarrow \FZ_1(\CS)$. In the fermionic cases, $n\Rep(G)$ is replaced by $n\Rep(G,z)$.

\end{expl}

\begin{rem} \label{rem:lqs+skeleton}
The same topological skeleton ${}^{\FZ_1(\CS)}\CS$ can be associated to different gapped/gapless phases depending on what local quantum symmetry we assign. We discuss a few examples for a fusion 1-category $\CS$. 
\bnu
\item Given two different unitary rational VOA's $V$ and $V'$ with non-trivial central charges and two braided equivalences $\phi: \Rep(V) \to \FZ_1(\CS)$ and $\phi': \Rep(V') \to \FZ_1(\CS)$. Then the triples $(V,\phi,{}^{\FZ_1(\CS)}\CS)$ and $(V',\phi',{}^{\FZ_1(\CS)}\CS)$ describe two different anomalous 1d gapless phases. Any two holomorphic VOA's are examples of such pairs of VOA's. One can also choose $V'$ to be a modular invariant closed CFT (i.e. a rational full field algebra). Then the triple $(V',\phi',{}^{\FZ_1(\CS)}\CS)$ describes an anomaly-free 1d gapless phase \cite{KZ21}. 

\item Let $\CS=\Rep(G)$ for a finite group $G$. If we identify $\FZ_1(\CS)$ with $\Rep(V^G)$ by a braided equivalence $\phi: \Rep(V^G) \to \FZ_1(\CS)$, where $V^G$ is the $G$-invariant sub-VOA of a holomorphic VOA $V$ (assuming the folklore conjecture \cite{DVVV89,Kir02,DNR21}), we obtain a 1d anomalous gapless phase. If we associate $\FZ_1(\CS)$ to an onsite symmetry $G$, which can be viewed as a proper orbifold theory\footnote{In \cite{KZ22}, we provide an alternative mathematical theory of local quantum symmetries based on certain nets of local operators. 
} 
of the trivial VOA $\Cb$, we obtain a 1d gapped SPT order. Moreover, there is the braided equivalence $\phi: \FZ_1(\Rep(G)) \to \FZ_1(\CS)$ defining the SPT order (see \cite[Corollary 2.25, Remark 2.26]{KZ22}). For example, if $\phi\simeq \id_{\FZ_1(\Rep(G))}$, then it defines the trivial 1d SPT order; otherwise, it defines a non-trivial SPT order. 

\enu
\end{rem}

\begin{expl}
This enriched-category description of topological skeleton also works for symmetry-breaking phases. Let $n\vect_G$ be the category of $G$-graded $n$-vector spaces. Applying topological Wick rotations to Remark\,\ref{rem:SET+symmetry-breaking}, one can see that ${}^{\FZ_1(n\Rep(G))}n\vect_G$ should describe a (spontaneous) symmetry-breaking phase. In this work, we prove this fact explicitly for $n=1$ and $G=\Zb_2$. 
\end{expl}

The operators (or chiral/non-chiral fields) in $V$ should be viewed as symmetric local operators. An object in $\Rep(V)$ should be viewed as a topological sector of symmetric nonlocal operators, i.e. an invariant subspace of all symmetric operators under the action of all symmetric local operators. For example, the operators in $M_{a,b}$ are all non-local operators because they can only live at the end point of a non-trivial TDL unless both $a$ and $b$ are the trivial TDL. The mathematical theory of local quantum symmetries for general gapped/gapless quantum liquids is far beyond this work and is developed in \cite{KZ22}.

An object in $\CS$ is a TDL (or a topological excitation from a spatial point of view). In a lattice model realization of the phase, a TDL amounts to a topological sector of states in the total Hilbert space $\CH_{tot}$ of the lattice model. By a topological sector of states, we mean an invariant subspace of states in $\CH_{tot}$ under the action of symmetric local operators and the symmetries. For example, in the 2d toric code model realization of the 2d $\Zb_2$ topological order, the four particle-like excitations $\one,e,m,f$ are precisely given by four topological sectors of states. 

\medskip
In summary, the enriched fusion category ${}^{\FZ_1(\CS)}\CS$ summarizes all observables in spacetime for all 1d gapped/gapless phases with/without symmetries up to the local quantum symmetry. In a lattice model realization of a 1d gapped phase, we expect that
\bnu
\item objects in $\CS$ are the topological sectors of states in $\CH_{tot}$; 
\item objects in $\FZ_1(\CR)$ (or $\FZ_1(\CS)$) are the topological sectors of symmetric non-local operators. 
\enu

In this work, we check the validity of this proposal by rediscovering the topological skeleton ${}^{\FZ_1(\CS)}\CS$ from two concrete 1d lattice models: the Ising chain and the Kitaev chain.

\bigskip
\noindent {\bf Acknowledgments}: 
We would like to thank Chun-Yu Bai, Gang Chen, Xiao-Liang Qi, Chenjie Wang, Zheng-Yu Weng, Rongge Xu and Zhi-Hao Zhang for helpful discussion and comments. LK is supported by NSFC under Grant No. 11971219 and Guangdong Provincial Key Laboratory (Grant No.2019B121203002) and Guangdong Basic and Applied Basic Research Foundation under Grant No. 2020B1515120100. XGW is partially supported by NSF DMR-2022428 and by the Simons Collaboration on Ultra-Quantum Matter, which is a grant from the Simons Foundation (651440). HZ is supported by NSFC under Grant No. 11871078.

\section{Ising chain and Kitaev chain}

In Section\,\ref{sec:general-discussion}, we explain that a gapped quantum liquid can be described by the observables in the long wave length limit (\lwll). In Section\,\ref{sec:sectors}, for 1d gapped lattice models, we explain that there are two types of observables in \lwll: the topological sectors of operators and those of states, and together they form an enriched fusion category with a trivial monoidal center. In Section\,\ref{sec:Ising-chain} and \ref{sec:Kitaev-chain}, we show explicitly that observables in all gapped phases realized by the Ising chain and the Kitaev chain indeed form enriched fusion categories with trivial centers. We also show that the observables on the boundaries of these phases form enriched categories such that the boundary-bulk relation holds, i.e. the bulk is the center of a boundary \cite{KWZ15,KWZ17}. 
These results provide solid evidence of the proposal in \cite{KZ20b} that the enriched-categorical description works for all gapped symmetry-breaking phases, topological orders, SPT/SET orders and CFT-type gapless phases. All examples of enriched (fusion) categories that appear in this section are briefly reviewed in Appendix\,\ref{sec:enriched-categories}.

\subsection{Quantum phases and observables} \label{sec:general-discussion}
Landau's theory of phases and phase transitions is based on the idea of symmetry breaking. This theory was so successful that it  led to the wrong belief that Landau's theory works for all quantum phases until the discovery of exotic new phases beyond Landau's paradigm. In retrospect, we can see that Landau's theory was not developed from the first principle\footnote{The meaning of ``the first principle'' varies as we change our point of view. From a categorical point of view, a notion of a phase can be understood via its relation (i.e. domain walls or phase transitions) to all phases. The point of view taken here is a reductionist one, i.e. defining a phase by its microscopic realizations or its macroscopic observables.}, 
by which we mean first defining the notion of a phase, then finding a way to characterize a phase transition. Instead, Landau's theory was developed from the study of a concrete phase transition. The tools and the language developed from this study automatically provide a way to distinguish different phases by the so-called order parameters and symmetries. However, it does not provide a priori reason for the completeness of the characterization of a phase by its symmetries. 

\medskip
The discovery of new gapped quantum phases beyond Landau's paradigm (such as fractional quantum Hall states) provides us a chance and motivation to study the notion of a gapped quantum phase from the first principle. Indeed, it has already motivated many attempts to define the notion of a gapped quantum liquid precisely from both the microscopic perspective \cite{CGW10a,ZW15} and the macroscopic perspective (see for example \cite{Kit06,KW14,KWZ15,LKW18,LW19,KTZho20,JF20,KLWZZ20a,KZ20b}). 

First, since a gapped quantum liquid can be realized by lattice models, there should be a microscopic definition based on lattice models. More precisely, a gapped quantum liquid phase should be defined as an equivalence class of lattice models. The general idea of the equivalence relation between two models is a path connecting two models in the space of models without closing the gap and without changing the ground state degeneracy anywhere on the path. More precisely, since only the ground state is physically relevant at zero temperature, in two interesting attempts \cite{CGW10a,ZW15}, the equivalence relation was defined directly for the ground state by local unitary transformations and the stacking of the product states. This is, however, not the final word about the equivalence relation. The real challenge lies in how to formulate it precisely and prove its compatibility with the macroscopic definition\footnote{As far as we know, there is no work on how to connect microscopic definition with a macroscopic one. For example, it is not clear or even puzzling that two lattice models defined at different RG fixed points realize the same quantum phase, such as the Levin-Wen models \cite{LW05}, can be connected by a path without closing the gap nor changing GSD. One way out is to add local quantum symmetry to the description of a topological order \cite{KZ20b} (see also Section\,\ref{sec:outlooks}).}.

Secondly, the notion of a quantum phase is defined at the thermodynamics limit and at zero temperature. At zero temperature, regardless gapped or gapless, only physically relevant observables are those survived in \lwll. Therefore, a quantum phase should be described by all observables (in \lwll) of a family of lattice models connected by small  symmetry-allowed perturbations. A careful analysis of all observables in a lattice model should lead us to such a description. Indeed, this analysis was done for many lattice models, such as the quantum double models \cite{Kit03} and the Levin-Wen models \cite{LW05}, and led to the correct categorical descriptions of 2d topological orders \cite{Kit03,KK12}. Ironically, this analysis has never been carried out for symmetry-breaking phases within Landau's paradigm. It turns out that this study is not so easy if you do not know what you are looking for. In this work, guided by the proposal in \cite{KZ20b} that 1d quantum phases should be described by enriched fusion categories (see Section\,\ref{sec:TWR}),
we start to do this long-overdue homework for two simple 1d lattice models: the Ising chain and the Kitaev chain.

\subsection{Topological sectors of operators and states} \label{sec:sectors}
For a given 1d lattice model with a total Hilbert space $\CH_{tot}=\otimes_{i\in\Zb} \CH_i$ and a Hamiltonian with only local interactions, many microscopic degrees of freedom are not observable in \lwll. For example, individual states in $\CH_{tot}$ and microscopic local operators are not observable in \lwll. It is similar to our daily experience. A physical object is always screened by the invisible cloud of microscopic degrees of freedom (or local operators) around the object. The observables in \lwll~are those that can be moved in and out of the cloud freely. They can come from non-local operators that are required to be {\it unconfined} (see Remark\,\ref{rem:unconfined}). For convenience, by a non-local operator we always mean a unconfined one unless it is declared otherwise. Moreover, from the \lwll~perspective, such a non-local operator is necessarily screened by local operators. Therefore, observables in \lwll~are not the individual non-local operators but the subspaces of non-local operators that are invariant under the action of local operators. Such an invariant subspace is called a {\it topological sector of operators} (see Remark\,\ref{rem:braiding}). The sector consisting of only local operators is denoted by $\one_\CB$. A morphism between two such sectors are operators that intertwine the action of local operators. We denote the category of the topological sectors of operators by $\CB$. This category $\CB$ has an obvious monoidal structure defined by the multiplication of operators. It turns out that $\CB$ also has a braiding structure (see Remark\,\ref{rem:braiding}). 
As a consequence, we expect the category $\CB$ to be a braided fusion category. Similar to the situation in 2+1D topological orders \cite{Kit03,KW14}, we expect that all sectors of operators can detect themselves via double braidings, or equivalently, the braidings of $\CB$ should be non-degenerate. 

\begin{rem} \label{rem:unconfined}
In this work, for convenience, we can treat a non-local operator as an infinitely long string of operators. Such a string of operators is called {\it unconfined} if the string remain tensionless under all symmetry allowed perturbations; it is called {\it confined} otherwise. 
\end{rem}

\begin{rem} \label{rem:braiding}
A rigorous study of topological sectors of operators is beyond this work, and is given in \cite{KZ22}. In a nutshell, ``local operators'' should be replaced by the net of local operators as in algebraic quantum field theories (see a review \cite{Haa92} and references therein). Then a topological sector of operators indeed becomes a sector of the net. The fusion and braiding structures of $\CB$ are defined in \cite{KZ22}. Now we provide some intuition about the braiding structure on $\CB$. The braiding structure on $\CB$ is encoded by operators living in 2D spacetime. It is different from that of anyons (or defect lines) defined in 2+1D spacetime. Two non-local operators $x$ and $y$ in 2D spacetime can be ``braided'' in the following sense: 
$$
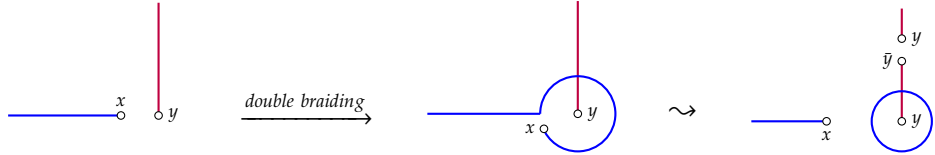
\raisebox{0.8em}{\begin{tikzpicture}
\draw [purple, thick] (0.5,0) -- (0.5,1.5) ;
\draw [blue,thick] (-1.5,0) -- (0,0) ;
\draw [fill=white] (0,0) circle [radius =0.05] ;
\draw [fill=white] (0.5, 0) circle [radius =0.05] ;
\node [above] at (0, 0) {\scriptsize $x$};
\node [right] at (0.5, 0) {\scriptsize $y$};
\end{tikzpicture}}
\quad\quad\raisebox{1em}{$\xrightarrow{double~braiding}$} \quad\quad
\begin{tikzpicture}
\draw [purple, thick] (0.5,0) -- (0.5,1.5) ;
\draw [blue, thick] (0,0) arc [radius=0.5, start angle=180, end angle= -160];
\draw [blue,thick] (-1.5,0) -- (0,0) ;
\draw [fill=white] (0.05,-0.2) circle [radius =0.05] ;
\draw [fill=white] (0.5, 0) circle [radius =0.05] ;
\node [left] at (0.05, -0.2) {\scriptsize $x$};
\node [right] at (0.5, 0) {\scriptsize $y$};
\end{tikzpicture}
\quad\quad\raisebox{1.3em}{$\leadsto$}\quad\quad
\begin{tikzpicture}
\draw [purple, thick] (0.5,1.1) -- (0.5,1.5) ;
\draw [purple, thick] (0.5,0) -- (0.5,0.8) ;
\draw [fill=white] (0.5, 0.8) circle [radius =0.05] ;
\draw [blue, thick] (0.5,0) circle [radius=0.4];
\draw [blue,thick] (-1.5,0) -- (-0.5,0) ;
\draw [fill=white] (-0.5,0) circle [radius =0.05] ;
\draw [fill=white] (0.5, 0) circle [radius =0.05] ;
\draw [fill=white] (0.5, 1.1) circle [radius =0.05] ;
\node [right] at (0.5, 1.1) {\scriptsize $y$};
\node [left] at (0.5, 0.8) {\scriptsize $\bar{y}$};
\node [below] at (-0.5, 0) {\scriptsize $x$};
\node [right] at (0.5, 0) {\scriptsize $y$};
\end{tikzpicture}
$$
In the second step $\leadsto$, we introduce a local operator that creates a pair $(y,\bar{y})$ such that the purple line breaks into two parts. The $\bar{y}y$ part becomes a local operator. This is possible because $\CB$ has duals (a natural physical requirement). Similarly, we introduce a local operator that creates a pair $(x,\bar{x})$ near $x$, then annihilates the $\bar{x}$ with the original $x$, we obtain the third picture. Comparing the third picture with the first one, we see an additional local operator $\bar{y}\bar{x}xy\bar{x}x$. By choosing $x,\bar{x},y,\bar{y}$ properly, this local operator can encode the information of the double braiding of the two sectors associated to $x$ and $y$ (see Section\,\ref{sec:J=0} for an example). 
\end{rem}

There is another type of observables, which are called topological excitations from a spatial perspective, or equivalently, topological defect lines (TDL) from a spacetime perspective. It is well known from the lattice model realizations of 2+1D topological orders, such as the toric code model \cite{Kit03}, a TDL (or an anyon) can be defined as a {\it topological sector (or superselection sectors) of states}, which is defined to be a subspace of $\CH_{tot}$ that are invariant under the action of local operators. The sector containing the vacuum is called the vacuum sector denoted by $\one_\CS$. The 1+1D cases are entirely the same. 

It is clear that the topological sectors of operators act on those of states. We denote the space of operators mapping a sector of states $a$ to another sector $b$ by $\hom(a,b)$, then $\hom(a,b)$ can be viewed as an object in $\CB$. The set of all sectors of states, together with the spaces of morphisms $\hom(a,b)$, form a category $\CS^\sharp$ enriched in $\CB$. A portrait of these $\hom(a,b)$ as observables on the 1+1D world sheet is given in Figure\,\ref{fig:observables} with $M_{a,b}$ representing $\hom(a,b)$. If we replace the hom space $\hom(a,b)$ in $\CS^\sharp$ by a vector space $\hom_\CS(a,b):=\hom_\CB(\one, \hom(a,b))$, we obtain an ordinary category $\CS$, which is reasonable to be called the category of the topological sectors of states (or TDL's or topological excitations).

Both $\CS^\sharp$ and $\CS$ are equipped with fusion products because TDL's can be fused (horizontally in Figure\,\ref{fig:observables}). The vacuum sector $\one_\CS$ plays the role of the tensor unit. Moreover, the fusion of two sectors of states should be compatible with the fusion of operators that can create these two sectors of states from the vacuum. This compatibility is rather complicated but can be mathematically summarized by the condition that $\CS$ is equipped with a braided monoidal functor $\phi: \CB \to \FZ_1(\CS)$, where $\FZ_1(\CS)$ is the Drinfeld center of $\CS$ \cite{KZ20b}. 
The braided monoidal functor $\phi$ provides a canonical construction of a $\CB$-enriched fusion category ${}^\CB\CS$ \cite{MP19}. It is natural to expect that $\CS^\sharp = {}^\CB\CS$ as $\CB$-enriched fusion categories\footnote{A lengthy proof of this fact in the case of 1+1D CFT's was given in \cite[Section\, 6]{KZ20a}. We expect that a similar proof works for gapped 1d phases (with symmetries) with the vertex operator algebra in \cite[Section\, 6]{KZ20a} replaced by a more general local quantum symmetry, which is clarified in \cite{KZ22}.}. Note that, when we choose to use $ {}^\CB\CS$ instead of ${}^{\FZ_1(\CS)}\CS$, we have already include some information of the local quantum symmetry (see Section\,\ref{sec:TWR}). By the boundary-bulk relation \cite{KWZ15,KWZ17}, the enriched fusion category ${}^\CB\CS$ describes an anomaly-free 1d phase if and only if $\FZ_1({}^\CB\CS)=\vect$. 

\void{
Different from 2+1D, where anyons can be detected by double braidings in the spatial dimension, in 1+1D, there is no braiding in the spatial dimension. If the 1+1D phase is anomaly-free, then the TDL's must be detectable by observables in the spacetime. We impose the following two anomaly-free conditions.  
\bnu
\item All topological sectors of states can be created from the vacuum by non-local operators.  
\item Topological sectors of states (or excitations) can be detected by the non-local operators. 
\enu
These conditions amount to a simple mathematical condition: the functor $\phi$ is an equivalence. Moreover, $\phi$ provides a canonical construction of a $\CB$-enriched fusion category ${}^\CB\CS$ \cite{MP19}. It is natural to expect that $\CS^\sharp \simeq {}^\CB\CS$ as $\CB$-enriched fusion categories\footnote{A lengthy proof of this fact in the case of 1+1D CFT's was given in \cite[Section\, 6]{KZ20a}. We expect that a similar proof works for gapped 1d phases (with symmetries) with the vertex operator algebra in \cite[Section\, 6]{KZ20a} replaced by a more general local quantum symmetry, which will be clarified elsewhere.}. Then the anomaly-free condition (i.e. $\phi$ is an equivalence) is equivalent to the usual bulk-is-trivial condition: $\FZ_1({}^\CB\CS)=\vect$ \cite[Corollary.\ 5.4]{KZ18}. 
}

\begin{rem} \label{rem:condensation}
If the 1d phase is anomaly-free, i.e. $\FZ_1({}^\CB\CS)=\vect$, then the vacuum sector $\one_\CS$ of states provides a condensation of $\CB$ \cite{KZ18}. More precisely, $A:=\hom_{{}^\CB\CS}(\one_\CS,\one_\CS)$ defines a Lagrangian algebra in $\CB$ and is condensed on the the vacuum sector $\one_\CS$ of states \cite{Kon14}. Moreover, $\CS$ can be recovered from $A$ as the category $\CB_A$ of right $A$-modules in $\CB$. This condensation interpretation is rather convenient for later studies. 
\end{rem}

When we do not impose any symmetry, all non-local operators are confined by introducing arbitrary perturbations. In other words, without imposing any symmetry, we obtain $\CB=\vect$. If the phase is anomaly-free, then it is necessary that $\CS=\vect$. This is just another way to see that there is no non-trivial anomaly-free 1d topological order. Moreover, we also recover the fact that an anomalous 1d topological order can be described by a fusion category $\CS$.

If we impose an onsite symmetry given by a finite group $G$, then all the small perturbations are required to respect the symmetry. In this case, the term ``local operators'' needs to be replaced by ``symmetric local operators'', and the topological sectors of operators need to be replaced by those of symmetric operators. More precisely, in the presence of an onsite symmetry, we again have two categories $\CB$ and $\CS$: 
\bnu
\item an object in $\CS$ is a TDL (or a particle-like topological excitation) or a topological sector of states, which is defined by an invariant subspace of $\CH_{tot}$ under the action of symmetric local operators and the symmetries. 

\item an object in $\CB$ is a topological sector of symmetric operators, which is defined by a subspace of all (potentially non-local) operators invariant under the action of symmetric local operators and the symmetries (see Remark\,\ref{rem:rigorous-result}). 
\enu
Altogether they form an enriched fusion category ${}^\CB\CS$ defined by a braided equivalence $\phi: \CB \to \FZ_1(\CS)$. We demonstrate this picture in later subsections through concrete 1d lattice models.

\subsection{Ising chain} \label{sec:Ising-chain}

Consider a 1d Ising chain\footnote{Note that $\otimes_{i\in\Zb} \Cb_i^2$ is not mathematically well-defined but should be viewed as a proper $N\to \infty$ limit of $\otimes_{-N<i<N} \Cb_i^2$ spanned by finite energy states.}: $\CH_{tot}=\otimes_{i\in\Zb} \Cb_i^2$ with the Hamiltonian defined as follows: 
\[
H = - \sum_i B X_i - \sum_i J Z_i Z_{i+1},
\]
where $X_i$ and $Z_i$ are Pauli matrices
$$
X = \left( \begin{array}{cc} 0 & 1 \\ 1 & 0 \end{array} \right), \quad\quad
Z= \left( \begin{array}{cc} 1 & 0 \\ 0 & -1 \end{array} \right).
$$ 
We set $|0\rangle_i, |1\rangle_i \in \Cb_i$ to be the eigenstates of $Z_i$, i.e. $Z_i|0\rangle_i=|0\rangle_i$ and $Z_i|1\rangle_i=-|1\rangle_i$, and set
$$
|+\rangle_i = \frac{1}{\sqrt{2}}(|0\rangle_i + |1\rangle_i), \quad\quad |-\rangle_i = \frac{1}{\sqrt{2}}(|0\rangle_i - |1\rangle_i).
$$
It is clear that $X_i|\pm\rangle_i=\pm|\pm\rangle_i$. 

\subsubsection{The $J=0$ case} \label{sec:J=0}
Now we consider the case $J=0$ and $B\approx 1$. In this case, the ground state is 
$$
|\Omega\rangle=|\cdots ++++\cdots\rangle. 
$$
The system is gapped. Note that $X_i$ is a local $\Zb_2$ symmetry, and $U = \otimes_i X_i$ defines a global onsite $\Zb_2$ symmetry.

\medskip
If we do not impose any symmetry, the only topological sector of states is the vacuum sector, denoted by $\one$. The only topological sector of operators is the trivial one. Indeed, in this case, all non-local operators can be confined by adding proper small perturbations. For example, the operators 
$$
m_i = \otimes_{k\leq i} X_k \quad \mbox{and} \quad Um_j 
$$ 
are confined by adding the term $-\sum_i KZ_i$ to the Hamiltonian. As a consequence, when $J=0$, the phase is the trivial 1d topological order and can be mathematically described by the category $\hilb$ of finite dimensional vector spaces, which has a unique simple object $\one$.

\medskip
Now we impose the $U$-symmetry. The ground state $|\Omega\rangle$ preserves the $U$-symmetry. We call an operator $P$ preserving the $U$-symmetry (i.e. $[P,U]=0$) a {\it $U$-symmetric operator}. For example, both the identity operator $1$ and $Z_iZ_{i+1}$ are $U$-symmetric local operators, and $m_j$ is a $U$-symmetric non-local operator. Although the operator $Z_i$ breaks the $U$-symmetry as a local operator, it can be viewed as a $U$-symmetric non-local operator because $Z_i = \otimes_{k\geq i}(Z_kZ_{k+1})$ (see Remark\,\ref{rem:two-Zi}). Moreover, $Z_i$ and $m_j$ are unconfined by any $U$-symmetric perturbations of the Hamiltonian. 

\begin{rem} \label{rem:two-Zi}
We set $Z_{i,j}:=\otimes_{i\leq k \leq j} Z_kZ_{k+1}$. Strictly speaking, $Z_i$ is not the same as $Z_{i,\infty}$ because the later has another $Z_j$ at $j \approx \infty$. However, $Z_i$ catches all the corrected local properties of $Z_{i,\infty}$ near the site $i$. So it is harmless and convenient to apply this identification $Z_i = Z_{i,\infty}$. Alternatively, one can use the string operator $Z_{i,j}$ with the string length $|j-i|$ much longer than the given characteristic length (or simply $Z_{i,\infty}$). The final result is irrelevant to the choice. 
\end{rem}

Now the total Hilbert space splits into two topological sectors of states labeled by symmetry charges. We denote the sector associated to the vacuum $|\Omega\rangle$ by $\one$, and the sector associated to the non-trivial symmetry charges by $e$. The trivial sector $\one$ is viewed as a trivial particle or a $\one$-particle. The lowest energy states in the sector $e$ are
$$
|e\rangle_i:=Z_i |\Omega\rangle = | \cdots ++ -_i +++ \cdots \rangle, \quad\quad \forall i\in\Zb,
$$
each of which represents an $e$-particle located at site $i$. 
The following state
$$
Z_iZ_j |\Omega\rangle = | \cdots ++ -_i +++ -_j +++ \cdots\rangle 
$$
represents two $e$-particles located at site $i$ and site $j$. This immediately implies the following fusion rules: $\one \otimes e = e\otimes \one =e,\, e \otimes e =\one$, 
which coincide with the fusion rules in the category $\Rep(\Zb_2)$ of $\Zb_2$-representations.


A topological sector of $U$-symmetric operators is invariant under the action of $U$-symmetric local operators and the symmetries. For example, all $U$-symmetric local operators and $U$ are in the trivial sector; and $Y_k, Z_k$ belong to the same topological sector because $Y_k = -i Z_kX_k$. The operator $m_i$ is a $U$-symmetric non-local operator. Although $m_i$ does not create a new particle from $|\Omega\rangle$, it plays a non-trivial role in the model. 

 By abusing the notation, we denote the topological sectors associated to the $U$-symmetric operators $1, m_i, Z_j, m_iZ_j$ by $\one, m, e, f$, respectively. For $x,y=\one,e$, we denote the space of $U$-symmetric operators that map from $x$ to $y$ by $\hom_{bulk}^{J=0}(x,y)$. Then we immediately obtain
\be \label{eq:internal-hom-J=0}
\hom_{bulk}^{J=0}(\one,\one) = \one \oplus m, \quad\quad \hom_{bulk}^{J=0}(\one,e)= \hom_{bulk}^{J=0}(e,\one)=e \oplus f, \quad\quad \hom_{bulk}^{J=0}(e,e)=\one \oplus m. 
\ee

Note that we have given the topological sectors associated to the operators $1, m_i, Z_i,m_iZ_j$ the same notations as those of anyons in the 2d toric code model or the simple objects in $\FZ_1(\Rep(\Zb_2))$ (see Appendix\,\ref{sec:enriched-categories}) because these topological sectors of operators provide a physical realization of the unitary module tensor category $\FZ_1(\Rep(\Zb_2))$ (recall Remark\ \ref{rem:rigorous-result}). Indeed, first note that these topological sectors of $U$-symmetric operators automatically satisfy the same fusion rules (defined by multiplying operators) as those of $\FZ_1(\Rep(\Zb_2))$. Moreover, they also recover the braidings in $\FZ_1(\Rep(\Zb_2))$. For example, one can recover the double braiding between $e$ and $m$ in $\FZ_1(\Rep(\Zb_2))$ by first creating a pair of ``$m$-particles'' at site $i$ and $j$ for $i<j$ (by applying $m_im_j$ to $|\Omega\rangle$), then applying $Z_k$ for $i<k<j$, then annihilating two $m$-particles, then annihilating $Z_k$, one obtains $Z_km_im_jZ_km_im_j=-1$, which is precisely the double braiding between $e$ and $m$ in $\FZ_1(\Rep(\Zb_2))$ (recall Remark\,\ref{rem:braiding}). One can recover the double braiding between $m$ and $e$ by $m_kZ_iZ_jm_kZ_iZ_j=-1$.  


\begin{rem} \label{rem:rigorous-result}
Similar discussion of these operators and their relation to $\FZ_1(\Rep(\Zb_2))$ have already appeared in \cite{JW20}, where these operators were called {\it patch symmetry operators}, and $\FZ_1(\Rep(\Zb_2))$ was called the {\it categorical symmetry} and interpreted as the bulk of a boundary, which has the algebraic higher symmetry $\Rep(\Zb_2)$. The point of view taken in \cite{JW20} is different from but connected to ours precisely through a topological Wick rotation (see Section\,\ref{sec:TWR}). 

In \cite{KZ22}, the symmetric local operators are reformulated precisely as the topological net of symmetric local operators, and a topological sector of operators is defined precisely as a sector of this topological net. For a 1d gapped quantum system with a finite onsite symmetry $G$, the category of all the sectors of the associated topological net is proved rigorously to be $\FZ_1(\Rep(G))$ (see Appendix\,\ref{sec:enriched-categories}) \cite[Corollary\ 2.25, Remark\ 2.26]{KZ22}. 
\end{rem}

Comparing (\ref{eq:internal-hom-J=0}) with (\ref{eq:Z(RepZ2)-RepZ2}), we obtain our first main result. 
\begin{pthm} \label{pthm:J=0-bulk}
The Ising chain when $J=0, B\approx 1$ realizes the trivial 1d $\Zb_2$ SPT order, which can be described mathematically by the enriched fusion category ${}^{\FZ_1(\Rep(\Zb_2))}\Rep(\Zb_2)$. 
\end{pthm}

\begin{rem} \label{rem:ignore-composition}
Strictly speaking, we have to check the identity morphisms (\ref{eq:identity}), the compositions of morphisms (\ref{eq:comp-1})-(\ref{eq:comp-2}) and the horizontal fusion morphisms, such as (\ref{eq:horizontal-fusion}), before we make the claim in Theorem${}^{\mathrm{ph}}$\,\ref{pthm:J=0-bulk}. Since these defining structures of enriched fusion categories are mathematically technical, and checking their coincidence with lattice models is straightforward and rather trivial in the $\Zb_2$-symmetry case, we decide to leave this checking as an exercise for all cases in this work. But for a non-abelian onsite symmetry $G$, this exercise can be non-trivial and interesting. 
\end{rem}

\begin{rem} \label{rem:cat-symmetry}
The categorical symmetry $\FZ_1(\Rep(\Zb_2))$ does not depend on the Hamiltonian. It only depends on the symmetry. This justifies the proposal that the bulk $\FZ_1(\Rep(\Zb_2))$ of the anomalous 1d phase defined by $\Rep(\Zb_2)$ should be viewed as the {\it categorical symmetry} of the 1d phase \cite{JW20,KLWZZ20b} (see also Remark\,\ref{rem:terminology}). Moreover, $\Rep(\Zb_2)$ is called the {\it algebraic higher symmetry} of the 1d phase \cite{JW20,KLWZZ20b}. 
\end{rem}

\begin{rem} \label{rem:condensation-1}
The trivial action of $m_i$ on $|\Omega\rangle$ can be interpreted as a condensation of the ``$m$-particles'' (or equivalently, the Lagrangian algebra $\one\oplus m$ \cite{Kon14}) in the categorical symmetry provided by the vacuum sector of states (recall Remark\,\ref{rem:condensation}). The multiplication of the Lagrangian algebra $\one\oplus m$ is given by (\ref{eq:comp-1}) and the unit is given by (\ref{eq:identity}). 
\end{rem}

\begin{rem} \label{rem:fix-anomaly-in-time}
Mathematically, the categorical description ${}^{\FZ_1(\Rep(\Zb_2))}\Rep(\Zb_2)$ is anomaly-free in the sense that this enriched fusion category ${}^{\FZ_1(\Rep(\Zb_2))}\Rep(\Zb_2)$ has a trivial monoidal center \cite{KZ18}, i.e.
\be \label{eq:Z1=trivial}
\FZ_1({}^{\FZ_1(\Rep(\Zb_2))}\Rep(\Zb_2)) = \vect. 
\ee
It is worthwhile to compare the ``anomaly-fixing'' mechanics in this approach with that in the gauging-the-symmetry approach and that in the boundary-bulk-relation approach introduced in \cite{KLWZZ20a}. 
\bnu
\item[$(1)$] In the gauging-the-symmetry approach, since the category $\Rep(\Zb_2)$ of symmetry charges cannot be detected by the braidings, it 
was viewed in some sense as ``anomalous". The anomaly is fixed by the gauging process of introducing new particles. The total particles after gauging form a multi-fusion 1-category $\Fun(\Rep(\Zb_2),\Rep(\Zb_2))$, which has a trivial monoidal center \cite[Section\ 2.2.1]{KLWZZ20a}. 

\item[$(2)$] In the boundary-bulk-relation approach, the category $\Rep(\Zb_2)$ of symmetry charges is also viewed as anomalous, and the anomaly is fixed by the 1-dimensional higher bulk $\FZ_1(\Rep(\Zb_2))$. In particular, the $m$-particles in the bulk can detect the $e$-particles via the half-braidings, thus fixed the anomaly \cite[Section\ 3.2]{KLWZZ20a}. 

\item[$(3)$] What we have shown in this subsection is that once we impose the $U$-symmetry, the category of the topological sectors of symmetric operators is changed to $\FZ_1(\Rep(\Zb_2))$. By replacing $\Rep(\Zb_2)$ with ${}^{\FZ_1(\Rep(\Zb_2))}\Rep(\Zb_2)$, we fix the ``anomaly'' in $\Rep(\Zb_2)$ by operators in 1+1D spacetime in the sense that an $e$-particle is now detectable by a $U$-symmetric local operator $m_im_j$ as explained in the paragraph below Eq. (\ref{eq:internal-hom-J=0}). 
\enu
Notice that (2) and (3) are essentially equivalent if we apply topological Wick rotation (see Section\,\ref{sec:TWR}). Moreover, one can recover the category $\Fun(\Rep(\Zb_2),\Rep(\Zb_2))$ in (1) by closing the fan around the left-bottom corner of Picture (b) in Figure\,\ref{fig:J=0} as the consequence of the following identity: 
$$
\Fun(\Rep(\Zb_2),\Rep(\Zb_2)) = \Rep(\Zb_2) \boxtimes_{\FZ_1(\Rep(\Zb_2))} \Rep(\Zb_2).
$$ 
See \cite[Eq.\ (3.4)]{KWZ15} for more details, and see \cite{KLWZZ20a,KLWZZ20b} for more discussion of this closing-fan realization of gauging the symmetry.
\end{rem}

\subsubsection{Two gapped boundaries when $J=0$}

It is also natural to consider the model with a gapped boundary, i.e. $\CH_{tot} = \otimes_{i\geq 0} \Cb_i^2$. 

\medskip
If we do not impose any symmetry, there is only one possible boundary condition. The precise boundary Hamiltonian is irrelevant because there is no non-trivial particles and no non-trivial (unconfined) non-local operators. Therefore, the boundary phase can be described mathematically by the category $\hilb$. We illustrate this fact by the following picture. 
\be 
\begin{tikzpicture}[scale=0.6]
\draw[->-,ultra thick] (-3,1)--(3,1) node[midway,above] {\scriptsize $\hilb$} ;
\draw[fill=white] (-3.1,0.9) rectangle (-2.9,1.1) node[midway,above] {\scriptsize $\hilb$} ;
\draw[fill=white] (2.9,0.9) rectangle (3.1,1.1) node[midway,above] {\scriptsize $\hilb$} ;
\end{tikzpicture}
\ee

\medskip
If we impose the $U$-symmetry, there are two choices of boundary conditions. 

\bnu
\item {\bf $U$-symmetric boundary condition}: The precise boundary Hamiltonian is irrelevant as long as it preserves the $U$-symmetry. For convenience, we can choose the following boundary Hamiltonian that preserves the $U$-symmetry:
$$
H = - \sum_{i\geq 0} X_i. 
$$
In this case, there are still two topological excitations on the boundary: $\one$ and $e$. On the boundary $m_0$ becomes a $U$-symmetric local operator now, so are $U$ and $Um_0$ because the $U$-symmetry is preserved on the boundary. Hence, we obtain the following topological sectors of operators:  
\be \label{eq:symmetric_bdy_J=0}
\hom_{\mathrm{s-bdy}}^{J=0}(\one,\one)=\one, \quad\quad \hom_{\mathrm{s-bdy}}^{J=0}(\one,e)=\hom_{\mathrm{s-bdy}}^{J=0}(e,\one)=e, \quad\quad \hom_{\mathrm{s-bdy}}^{J=0}(e,e)=\one. 
\ee
Comparing (\ref{eq:symmetric_bdy_J=0}) with (\ref{eq:RepZ2-RepZ2}), we conclude that the boundary phase can be described by 
the $\Rep{\Zb_2}$-enriched 1-category ${}^{\Rep{\Zb_2}}\Rep{\Zb_2}$.

\item {\bf $U$-symmetry broken boundary condition}: For example, we can choose the following boundary Hamiltonian to break the $U$-symmetry only on the boundary: 
$$
H = -Z_0 - \sum_{i>0} X_i. 
$$
In this case, the ground state is $|1+++\cdots\rangle$, and $Z_0$ does not create a new sector of states from the vacuum. Or equivalently, we can say that $e$-particles condense on this boundary. Moreover, $Z_0$ becomes a local operator because the $U$-symmetry is broken on the boundary. So is $Z_{0,\infty}$ because $Z_0Z_{0,\infty}$ is now a local operator. Although $m_0$ becomes a local operator, $Um_0$ remains a non-local operator and defines a non-trivial topological sector of operators because the $U$-symmetry is broken on the boundary. As a consequence, we obtain 
\be \label{eq:sb_bdy_J=0}
\hom_{\mathrm{sb-bdy}}^{J=0}(\one,\one)=\one\oplus m.
\ee 
Comparing (\ref{eq:sb_bdy_J=0}) with (\ref{eq:HilbZ2-Hilb}), we see that the observables on 
the $U$-symmetry broken boundary form an enriched category ${}^{\hilb_{\Zb_2}}\hilb$. 

\enu

\begin{rem}
If we consider the boundaries on the right side, i.e. $\CH_{tot}=\otimes_{i\leq 0} \Cb_i^2$. The categorical descriptions of the boundaries remain the same. 
\end{rem}

\begin{rem}
It is clear that the observables in the bulk act on those on the boundary. Therefore, the categorical description of a boundary is necessarily a module over that of the bulk. Indeed, ${}^{\Rep{\Zb_2}}\Rep{\Zb_2}$ and ${}^{\hilb_{\Zb_2}}\hilb$ are both closed modules over ${}^{\FZ_1(\Rep(\Zb_2))}\Rep(\Zb_2)$ (see \cite[Definition\ 3.18]{KZ21}). 
By \cite[Corollary\, 4.39]{KYZZ21}, and the boundary-bulk relation (i.e. the bulk is the center of a boundary) \cite{KWZ17} holds for both boundaries, i.e. 
\be \label{eq:Z0=bulk}
\FZ_0({}^{\Rep{\Zb_2}}\Rep{\Zb_2})={}^{\FZ_1(\Rep(\Zb_2))}\Rep(\Zb_2) = 
\FZ_0({}^{\hilb_{\Zb_2}}\hilb), 
\ee
where $\FZ_0$ denotes the $E_0$-center of an enriched category \cite[Section\,4.4]{KYZZ21}. The identity (\ref{eq:Z1=trivial}) automatically follows from (\ref{eq:Z0=bulk}) by the fact that the center of a center is trivial \cite[Remark\,5.28]{KYZZ21}. This fact is the mathematical counterpart of the obvious physical fact that the bulk of a bulk is trivial. 
\end{rem}

\begin{figure}
\[
\begin{array}{cc}

\begin{tikzpicture}[scale=0.6]
\draw[->-,ultra thick] (-3,1)--(3,1) node[midway,below] {\scriptsize ${}^{\FZ_1(\Rep(\Zb_2))}\Rep(\Zb_2)$} ;
\draw[fill=white] (-3.1,0.9) rectangle (-2.9,1.1) node[midway,above] {\scriptsize ${}^{\Rep(\Zb_2)}\Rep(\Zb_2)$} ;
\draw[fill=white] (2.9,0.9) rectangle (3.1,1.1) node[midway,above] {\scriptsize ${}^{\hilb_{\Zb_2}}\hilb$} ;
\end{tikzpicture}

& 

\begin{tikzpicture}[scale=0.6]
\fill[gray!20] (-3,0) rectangle (3,3) ;
\draw[->-,very thick] (-3,0)--(3,0) node[midway,below] {\footnotesize $\Rep(\Zb_2)$} ;
\draw[->-,very thick] (-3,3)--(-3,0) node[midway,left] {\scriptsize $\Rep(\Zb_2)$} ;
\draw[->-,very thick] (3,3)--(3,0) node[midway,right,scale=0.8] {$\hilb_{\Zb_2}$} ;

\draw[fill=white] (-3.1,-0.1) rectangle (-2.9,0.1) node[midway,below] {\scriptsize $\Rep(\Zb_2)$} ;
\draw[fill=white] (2.9,-0.1) rectangle (3.1,0.1) node[midway,below] {\scriptsize $\hilb$} ;
\node[above] at (-0,1.3) {$\FZ_1(\Rep(\Zb_2))$} ;

\end{tikzpicture}

\\
(a)  & (b)
\end{array}
\]
\caption{These pictures illustrate two gapped boundaries of the trivial 1d $\Zb_2$ SPT order in two different ways. 
}
\label{fig:J=0}
\end{figure}
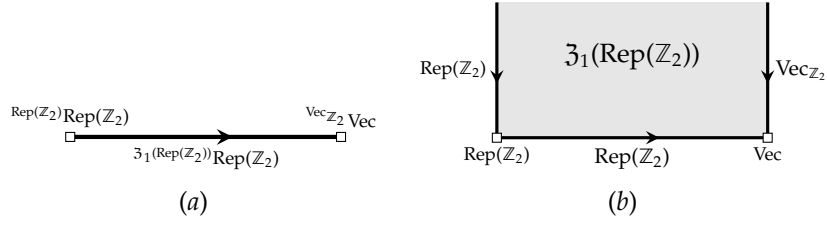

In Picture (a) of Figure\,\ref{fig:J=0}, we illustrate the 1d $\Zb_2$ SPT order, together with its two gapped boundaries that are constructed in this subsection. Picture (b) of Figure\,\ref{fig:J=0} depicts a 2d $\Zb_2$ topological order described by $\FZ_1(\Rep(\Zb_2))$, together with two gapped boundaries described by two fusion 1-categories $\Rep(\Zb_2)$ and $\hilb_{\Zb_2}$, respectively, and the trivial domain wall (defined by $\Rep(\Zb_2)$) in $\Rep(\Zb_2)$ and an invertible domain wall (defined by $\hilb$) between $\Rep(\Zb_2)$ and $\hilb_{\Zb_2}$. In particular, the vertical direction is the 2nd spatial direction. We see that Picture (a) can be obtained from Picture (b) by applying the topological Wick rotation \cite{KZ20a} (see also Section\,\ref{sec:TWR}).

\subsubsection{The $B=0$ case} \label{sec:B=0}
Now we consider the case $B=0$ and $J\approx 1$. we have $H=-\sum_{i} Z_iZ_{i+1}$. In this case, $U=\otimes_i X_i$ is still a global symmetry, but $X_i$ is not a local symmetry. The following two states
$$ 
|\cdots 000\cdots\rangle \quad\quad \mbox{and} \quad\quad |\cdots 111\cdots\rangle 
$$
are both ground states representing $U$-symmetry broken phases. 

\medskip
If we do not impose any symmetry and if we ignore perturbations, then total Hilbert space splits into four sectors $\CH_{ab}$ for $a,b=0,1$, where $\CH_{ab}$ is spanned by states $(\otimes_{k<i} |a\rangle_k)(\otimes_{k\geq i} |b\rangle_k)$ for $i\in\Zb$. We denote the topological sector associated to $\CH_{ab}$ by $s_{ab}$. Then we see immediately the fusion rules among them:
\be \label{eq:fusion-rule-s_ab}
s_{ab}\otimes s_{cd} = \delta_{bc}s_{ad}. 
\ee
Similar to the no-symmetry case when $J=0$, there are no non-local operators. Therefore, this phase is described mathematically by the unitary multi-fusion 1-category that consists of four simple objects $s_{00}, s_{01}, s_{10}, s_{11}$ satisfying the fusion rules (\ref{eq:fusion-rule-s_ab}). 
Mathematically, this multi-fusion category is precisely the category $\Fun(\hilb_{\Zb_2}, \hilb_{\Zb_2})$. However, this nice mathematical description requires fine tuning and is not stable under perturbations. By adding a small perturbation term say $-\sum_i K Z_i$ for $0<K\ll 1$, all sectors $s_{01},s_{10},s_{11}$ are gone. We obtain again the trivial phase described by $\hilb$. 

\begin{rem}
Although it needs fine tuning, the mathematical description $\Fun(\hilb_{\Zb_2}, \hilb_{\Zb_2})$ of a 1d bulk is natural and anomaly-free because the $E_1$-center (or Drinfeld center) of $\Fun(\hilb_{\Zb_2}, \hilb_{\Zb_2})$ is trivial, i.e. $\FZ_1(\Fun(\hilb_{\Zb_2}, \hilb_{\Zb_2}))=\hilb$. It naturally appears in the process of dimensional reductions of a 2d topological order \cite{KWZ15,AKZ17} (see Remark\,\ref{rem:dim-reduction-fine-tuning}). 
\end{rem}

Now we impose the $U$-symmetry. Note that none of $s_{ab}$ for $a,b=0,1$ are $U$-symmetric because the ground states break the symmetry. They form two $U$-symmetric topological sectors of states: 
$$
\one := s_{00}\oplus s_{11}, \quad\quad m= s_{01}\oplus s_{10},
$$ 
The fusion rules are $\one \otimes m =m \otimes \one=m$ and $m \otimes m=\one$, coinciding with those in $\hilb_{\Zb_2}$. 

\begin{rem}
Although the states in the sector $s_{00}\oplus s_{11}$ (such as $|\cdots 000 \cdots\rangle \pm |\cdots 111\cdots\rangle$) can carry different ``$U$-charges'', it is physically meaningless to split $s_{00}\oplus s_{11}$ further into two sectors according to the ``$U$-charges'' because the relative phase factors in the superposition of two states in two different superselection sectors are meaningless according to \cite{WWW52}.\footnote{This fact is compatible with the fact that the difficulty of creating a Schr\"{o}dinger Cat state grows exponentially (i.e. $\sim 2^N$) as the number of qubits $N$ approach $\infty$ (see a discussion of this fact from a modern perspective \cite{QR20}).} The sector $m$ is similar. 
\end{rem}


The $U$-symmetric non-local operator $Z_i$ (or rather $Z_{i,\infty}$ recall Remark\,\ref{rem:two-Zi}) acts on the vacuum $|\cdots 000 \cdots\rangle $ trivially. Using the same analysis as in Section\,\ref{sec:J=0}, we immediately obtain
$$ 
\hom_{bulk}^{B=0}(\one,\one) = \one \oplus e, \quad \hom_{bulk}^{B=0}(\one,m)=\hom_{bulk}^{B=0}(m,\one)=m \oplus f, \quad \hom_{bulk}^{B=0}(m,m)=\one\oplus e. 
$$
Comparing them with (\ref{eq:Z(RepZ2)-HilbZ2}), we obtain the following result (recall Remark\,\ref{rem:ignore-composition}). 
\begin{pthm}
The Ising chain when $B=0$ and $J\approx 1$ with the $U$-symmetry realizes a spontaneous symmetry-breaking phase, which can be described mathematically by the enriched fusion category ${}^{\FZ_1(\Rep(\Zb_2))}\hilb_{\Zb_2}$. 
\end{pthm}

\begin{rem} \label{rem:condensation-2}
The trivial action of $Z_i$ on the vacuum can be interpreted as the condensation of the ``$e$-particles'' (or equivalently, the Lagrangian algebra $\one\oplus e$ \cite{Kon14}) in the categorical symmetry provided by the vacuum sector of states (recall Remark\,\ref{rem:condensation}). 
\end{rem}

\begin{rem}
Note that $\Rep(\Zb_2)=\vect_{\Zb_2}$ as fusion categories. We can identify ${}^{\FZ_1(\Rep(\Zb_2))}\hilb_{\Zb_2}$ with ${}_{\quad\quad m \leftrightarrow e}^{\FZ_1(\Rep(\Zb_2))}\Rep(\Zb_2)$. The enrichment in ${}_{\quad\quad m \leftrightarrow e}^{\FZ_1(\Rep(\Zb_2))}\Rep(\Zb_2)$ is twisted by the braided auto-equivalence $(e\leftrightarrow m)$ of $\FZ_1(\Rep(\Zb_2))$. Note that this braided auto-equivalence does not preserve the symmetry charges $\Rep(\Zb_2)$ in $\FZ_1(\Rep(\Zb_2))$. This coincides with the fact that $H^2(\Zb_2,U(1))$ is trivial. 
\end{rem}

\subsubsection{Two gapped boundaries when $B=0$}

Now we consider the same model with a boundary on the left side, i.e. $\CH_{tot}=\oplus_{i\geq 0} \Cb_i^2$. 

\medskip
Without imposing any symmetry, and without fine tuning, there is only one sector of states associated to the lattice with a boundary. It consists of the lowest energy state $|000\cdots\rangle$. The same boundary condition can be imposed on the right side. 
We illustrate two side boundaries in the following picture: 
\be \label{pic:two-bdy-no-symmetry-bulk-2}
\begin{tikzpicture}[scale=0.6]
\draw[->-,ultra thick] (-3,1)--(3,1) node[midway,above] {\scriptsize $\hilb$} ;
\draw[fill=white] (-3.1,0.9) rectangle (-2.9,1.1) node[midway,above] {\scriptsize $\hilb$} ;
\draw[fill=white] (2.9,0.9) rectangle (3.1,1.1) node[midway,above] {\scriptsize $\hilb$} ;
\end{tikzpicture}
\ee

\begin{rem}
It is worthwhile to discuss the boundaries of the fine tuned bulk (consisting of four sectors $s_{ab}$ for $a,b=0.1$) without imposing any symmetry because they often occur in the process of dimensional reductions. There are only two non-trivial sectors $s_0'$ and $s_1'$ of states associated to the lattice with a boundary. The sector $s_0'$ consists of the state $|000\cdots\rangle$. The sector $s_1'$ consists of the state $|111\cdots\rangle$. There is no unconfined non-local operators. Therefore, the boundary phase can be described by the category $\hilb_{\Zb_2}$ (forgetting its monoidal structure). Moreover, we have the following fusion rules: 
$$
s_a' \otimes s_{bc} =\delta_{ab} s_c', \quad\quad \forall a,b,c=0,1. 
$$
This fusion rule endows the category $\hilb_{\Zb_2}$ with a structure of right $\Fun(\hilb_{\Zb_2},\hilb_{\Zb_2})$-module. 
Similarly, if we choose a boundary on the right side, i.e. $\CH_{tot}=\oplus_{i\leq 0} \Cb_i^2$. Again the boundary phase can be described by the 1-category $\hilb_{\Zb_2}$ with two simple objects $s_a''$ and the fusion rules: 
$$
s_{ab} \otimes s_c'' = \delta_{bc} s_a,  \quad\quad \forall a,b,c=0,1. 
$$
This fusion rule defines a structure of a left $\Fun(\hilb_{\Zb_2},\hilb_{\Zb_2})$-module on $\hilb_{\Zb_2}$. We illustrate them in the following picture: 
\be \label{pic:two-bdy-fine-tuning}
\begin{tikzpicture}[scale=0.6]
\draw[->-,ultra thick] (-3,1)--(3,1) node[midway,above] {\scriptsize $\Fun(\hilb_{\Zb_2},\hilb_{\Zb_2})$} ;
\draw[fill=white] (-3.1,0.9) rectangle (-2.9,1.1) node[midway,above] {\scriptsize $\hilb_{\Zb_2}$} ;
\draw[fill=white] (2.9,0.9) rectangle (3.1,1.1) node[midway,above] {\scriptsize $\hilb_{\Zb_2}$} ;
\end{tikzpicture}
\ee
\end{rem}

\medskip
If we impose the $U$-symmetry, there are two choices of boundary conditions. 

\bnu

\item  {\bf $U$-symmetric boundary condition}:
For example, we can choose the boundary Hamiltonian as $H= -\sum_{i\geq 0} Z_iZ_{i+1}$. 
Now $m$-particles condense on the boundary. Only surviving particle on the boundary is $\one$. Hence, the category of boundary particles is $\hilb$ (forgetting the monoidal structure). In the neighborhood of the boundary, $m_0$ and $Um_0$ become local $U$-symmetric operators. The operator $Z_0$ is still a $U$-symmetric non-local operator. Therefore, we obtain 
$$
\hom_{s-bdy}^{B=0}(\one,\one)=\one\oplus e. 
$$
Comparing it with (\ref{eq:RepZ2-Hilb}), we see that the observables on this gapped boundary form the enriched category ${}^{\Rep(\Zb_2)}\hilb$.

\item {\bf $U$-symmetry broken boundary condition}: For example, we can choose the boundary Hamiltonian as $H= - Z_0 - \sum_{i> 0} Z_iZ_{i+1}$. In this case, $U$-symmetry is broken on the boundary. There are still two sectors of states consisting of 
$$
|0000\cdots\rangle
\quad\quad \mbox{and} \quad\quad Um_0|0000\cdots\rangle=|0111\cdots \rangle,
$$ 
respectively. Note that a bulk $m$-particle acts on the two boundary sectors of states as a non-trivial permutation. Hence, the boundary particles form the category $\hilb_{\Zb_2}$ (forgetting the monoidal structure). 
The operator $Z_0$ is now a local operator. The only non-trivial sector of operators consists of $U$ (because the symmetry is broken on the boundary). Therefore, we obtain 
$$
\hom_{sb-bdy}^{B=0}(\one,\one) = \hom_{sb-bdy}^{B=0}(m,m) = \one, 
\quad\quad
\hom_{sb-bdy}^{B=0}(\one,m) = \hom_{sb-bdy}^{B=0}(m,\one) = m. 
$$
Comparing them with (\ref{eq:HilbZ2-HilbZ2}), we see that the observables on this gapped boundary form the enriched category ${}^{\hilb_{\Zb_2}}\hilb_{\Zb_2}$. 

\enu

\begin{rem}
By \cite[Corollary\, 4.39]{KYZZ21}, the boundary-bulk relation still holds, i.e. 
$$
\FZ_0({}^{\Rep(\Zb_2)}\hilb)={}^{\FZ_1(\Rep(\Zb_2))}\hilb_{\Zb_2} = \FZ_0({}^{\hilb_{\Zb_2}}\hilb_{\Zb_2}).
$$ 
\end{rem}

\begin{figure}
\[
\begin{array}{cc}

\begin{tikzpicture}[scale=0.6]
\draw[->-,ultra thick] (-3,1)--(3,1) node[midway,below] {\scriptsize ${}^{\FZ_1(\Rep(\Zb_2))}\hilb_{\Zb_2}$} ;
\draw[fill=white] (-3.1,0.9) rectangle (-2.9,1.1) node[midway,above] {\scriptsize ${}^{\Rep(\Zb_2)}\hilb$} ;
\draw[fill=white] (2.9,0.9) rectangle (3.1,1.1) node[midway,above] {\scriptsize ${}^{\hilb_{\Zb_2}}\hilb_{\Zb_2}$} ;
\end{tikzpicture}

& 

\begin{tikzpicture}[scale=0.6]
\fill[gray!20] (-3,0) rectangle (3,3) ;
\draw[->-,very thick] (-3,0)--(3,0) node[midway,below] {\footnotesize $\hilb_{\Zb_2}$} ;
\draw[->-,very thick] (-3,3)--(-3,0) node[midway,left] {\scriptsize $\Rep(\Zb_2)$} ;
\draw[->-,very thick] (3,3)--(3,0) node[midway,right,scale=0.8] {$\hilb_{\Zb_2}$} ;

\draw[fill=white] (-3.1,-0.1) rectangle (-2.9,0.1) node[midway,below] {\scriptsize $\hilb$} ;
\draw[fill=white] (2.9,-0.1) rectangle (3.1,0.1) node[midway,below] {\scriptsize $\hilb_{\Zb_2}$} ;
\node[above] at (-0,1.3) {$\FZ_1(\Rep(\Zb_2))$} ;

\end{tikzpicture}

\\
(a)  & (b)
\end{array}
\]
\caption{These pictures illustrate two gapped boundaries of the $\Zb_2$-symmetry broken phase in two different ways. 
}
\label{fig:B=0}
\end{figure}
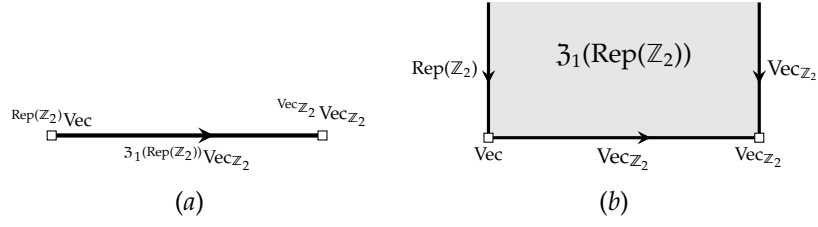

Figure\,\ref{fig:B=0} (a) illustrates the $\Zb_2$-symmetry broken 1d phase with two gapped boundaries. In Figure\,\ref{fig:B=0} (b), we depict a 2d topological order $\FZ_1(\Rep(\Zb_2))$, together with two different gapped boundaries $\Rep(\Zb_2)$ and $\hilb_{\Zb_2}$ and two 0d domain walls $\hilb_{\Zb_2}$ and $\hilb$. In particular, the vertical direction is the 2nd spatial direction. Again Figure\,\ref{fig:B=0} (a) can be obtained from Figure\,\ref{fig:B=0} (b) via a topological Wick rotation (see Section\,\ref{sec:TWR}). 

\begin{rem} \label{rem:dim-reduction-fine-tuning}
If we fuse the left vertical line in Figure\,\ref{fig:B=0} (b) with the horizontal line, we obtain a 1d bulk phase with a boundary as depicted in (\ref{pic:two-bdy-no-symmetry-bulk-2}); if we fuse the right vertical line with the horizontal line, we obtain a fine-tuned 1d bulk with a boundary as depicted in (\ref{pic:two-bdy-fine-tuning}). 
\end{rem}

\begin{rem}
When $J=B=1$, the system is at the critical point of a phase transition. We denote the Ising vertex operator algebra by $V_{\mathrm{Is}}$, its right moving counterpart by 
$\overline{V_{\mathrm{Is}}}$ and the Ising unitary modular tensor category by $\mathbf{Is}$, i.e. $\mathbf{Is}=\Mod_{V_{\mathrm{Is}}}$. Then this critical point (as a 1+1D gapless phase) might\footnote{There are other candidates. For example, one can replace $V_{\mathrm{Is}}\otimes_\Cb \overline{V_{\mathrm{Is}}}$ by the full field algebra $A=\one\boxtimes\one \oplus \psi \boxtimes \psi$, i.e. a condensation algebra in $\FZ_1(\mathbf{Is})$, we obtain another pair $(A, {}^{\FZ_1(\Rep(\Zb_2))}\Rep(\Zb_2)$. This is an interesting direction to explore.} be described by a pair 
$(V_{\mathrm{Is}}\otimes_\Cb \overline{V_{\mathrm{Is}}}, \, {}^{\FZ_1(\mathbf{Is})}\mathbf{Is})$ \cite{KZ21}. This also means that the enriched-category description works for both gapped and gapless phases. 
\end{rem}

\begin{rem}
One can also use a domain wall to connect two 1d gapped phases realized by the Ising chain by considering the following Hamiltonian 
$$
H= - \sum_{i<0} BX_i - \sum_{i\geq 0} J Z_iZ_{i+1}
$$ 
for $J = B \approx 1$. Note that both $m_{-1}$ and $Z_0$ acts trivially on the vacuum $|\cdots +++000 \cdots\rangle$. Similar to Remark\,\ref{rem:condensation-1} and \ref{rem:condensation-2}, all ``particles'' in the categorical symmetry are condensed by the vacuum sector $\one_{wall}$ of states on the wall, i.e. $\hom(\one_{wall},\one_{wall})=\one\oplus m\oplus e\oplus f$. Comparing it with (\ref{eq:ZZ2-vect}), it is clear that the wall can be described by the enriched category ${}^{\FZ_1(\Rep(\Zb_2))}\hilb$, which is an invertible ${}^{\FZ_1(\Rep(\Zb_2))}\Rep(\Zb_2)$-${}^{\FZ_1(\Rep(\Zb_2))}\hilb_{\Zb_2}$-bimodule. This implies that two different gapped phases realized in the Ising chain are Morita equivalent. Actually, the representation theory of enriched fusion categories predicts that there are other bimodules or domain walls. We illustrate one in the following picture: 
$$
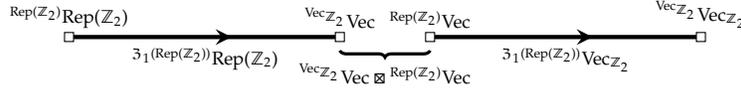
\begin{tikzpicture}[scale=0.6]
\draw[->-,ultra thick] (-3,1)--(3,1) node[midway,below] {\scriptsize ${}^{\FZ_1(\Rep(\Zb_2))}\Rep(\Zb_2)$} ;
\draw[fill=white] (-3.1,0.9) rectangle (-2.9,1.1) node[midway,above] {\scriptsize ${}^{\Rep(\Zb_2)}\Rep(\Zb_2)$} ;
\draw[fill=white] (2.9,0.9) rectangle (3.1,1.1) node[midway,above] {\scriptsize ${}^{\hilb_{\Zb_2}}\hilb$} ;
\draw[decorate,decoration=brace,very thick] (5,0.7)--(3,0.7) ;
\node at (4,0.2) {\scriptsize ${}^{\hilb_{\Zb_2}}\hilb\boxtimes{}^{\Rep(\Zb_2)}\hilb$} ;
\draw[->-,ultra thick] (5,1)--(11,1) node[midway,below] {\scriptsize ${}^{\FZ_1(\Rep(\Zb_2))}\hilb_{\Zb_2}$} ;
\draw[fill=white] (4.9,0.9) rectangle (5.1,1.1) node[midway,above] {\scriptsize ${}^{\Rep(\Zb_2)}\hilb$} ;
\draw[fill=white] (10.9,0.9) rectangle (11.1,1.1) node[midway,above] {\scriptsize ${}^{\hilb_{\Zb_2}}\hilb_{\Zb_2}$} ;
\end{tikzpicture}
$$
By the spatial equivalence introduced in \cite{Zhe17,KZ21}, all of this bimodules or domain walls are spatially equivalent due to the lack of thermodynamic limit of ``0d phases''. 
\end{rem}

\begin{rem}
Our restudy of the Ising chain explicitly shows that the enriched-category approach is capable of unifying the SPT orders with the (spontaneous) symmetry-breaking orders. 
\end{rem}

\subsection{1d Kitaev chain} \label{sec:Kitaev-chain}

Consider the Kitaev chain \cite{Kit03}. 
$$\CH = \bigotimes_j \CH_j,$$
$$
H = - \mu \sum_j c_j^\dagger c_j - t \sum_j (c_{j+1}^\dagger c_j + c_j^\dagger c_{j+1} ) + \Delta \sum_j (c_jc_{j+1} + c_{j+1}^\dagger c_j^\dagger),
$$
where $\CH_j$ is the super vector space of dimension $1|1$. We can rewrite the Hamiltonian by Majorana operators: 
$$
\gamma_{j,1} = c_j + c_j^\dagger, \quad \gamma_{j,2} = i(c_j-c_j^\dagger).
$$
We have $\gamma_{j,a}^\dagger = \gamma_{j,a}$ and $\{ \gamma_{j,a}, \gamma_{k,b}\}=2\delta_{jk} \delta_{ab}$ for $a,b=1,2$. Then we obtain
$$
H = \frac{\mu}{2} \sum_j (1-i\gamma_{j,1}\gamma_{j,2}) + \frac{t-\Delta}{2} \sum_j i \gamma_{j,1}\gamma_{j+1,2} - \frac{t+\Delta}{2} \sum_j i \gamma_{j,2}\gamma_{j+1,1}
$$
There is a fermion parity operator 
$$
U = \otimes_j \, i \gamma_{j,1}\gamma_{j,2}. 
$$


The $U$-symmetric local operators are generated by $\gamma_{j,1}\gamma_{j,2}$ and $\gamma_{j,1}\gamma_{j+1,2}$. There are $U$-symmetric nonlocal operators
$$
m_k = \prod_{j\leq k} i\gamma_{j,1}\gamma_{j,2}, \quad f_k = \prod_{j\leq k}\gamma_{j,2}\gamma_{j+1,2}, \quad e_k=m_kf_k. 
$$ 
There are again four simple sectors of symmetric operators $\one,m,f,e$ associated to the identity operator, $m_k, f_k, e_k$, respectively. The category of the topological sectors of operators form the braided fusion category $\FZ_1(\svec)$. As a braided fusion category, $\FZ_1(\svec)$ is braied equivalent to $\FZ_1(\Rep(\Zb_2))$, but the symmetry charges are embedded into $\FZ_1(\svec)$ according to $\one \mapsto \one, f\mapsto f$. 

It is helpful to compare this model with the Ising chain under the correspondence $X_j\mapsto i\gamma_{j,1}\gamma_{j,2}$ and $Z_j\mapsto\gamma_{j,2}$. However, different from $Z_j$, the operator $\gamma_{j,2}$ is a super operator. Thus the category $\Rep(\Zb_2)$ is generally replaced by the category $\svec$ of finite dimensional supervector spaces in this model. Note that $\Rep(\Zb_2)$ and $\svec$ are the same fusion category but different in their braidings. It is helpful to remind the readers of three different fusion subcategories of $\FZ_1(\Rep(\Zb_2))$ given in (\ref{eq:three-embedding}), which also explains our notations.


\subsubsection{The case $\mu=1$ and $t=\Delta=0$}

In the bulk, there are two topological sectors of states (or bulk excitations) labeled by symmetry charges $\one$ (the even parity) and $f$ (the odd parity) with the following obvious fusion rules: 
$$\one\otimes\one=f\otimes f=\one, \quad \one\otimes f=f\otimes\one=f.$$
Moreover, 
$$\hom_{bulk}^{kc1}(\one,\one) = \hom_{bulk}^{kc1}(f,f) = \one \oplus m,$$
$$\hom_{bulk}^{kc1}(\one,f) = \hom_{bulk}^{kc1}(f,\one) = f \oplus e.$$
Comparing them with (\ref{eq:Z1svec-svec}), we see that the observables in this 1d phase form the enriched fusion category ${}^{\FZ_1(\svec)}\svec$. According to 
Example\,\ref{expl:nd-SPT} (see also Remark\,\ref{rem:lqs+skeleton}), this phase is the trivial fermionic 1d SPT order with a $\Zb_2$ onsite symmetry. 
\medskip

When there is a boundary on the left defined by the same Hamiltonian restricting $i\ge0$, there are two boundary excitations $\one,f$, and we have 
$$\hom_{bdy}^{kc1}(\one,\one) = \hom_{bdy}^{kc1}(f,f) = \one,$$
$$\hom_{bdy}^{kc1}(\one,f) = \hom_{bdy}^{kc1}(f,\one) = f.$$
Comparing them with (\ref{eq:svec-svec}), we see that the observables on this boundary form the enriched category ${}^\svec\svec$. The 1d bulk phase and its boundary can be obtained from the 2d spatial configuration depicted in the left half of Figure\,\ref{fig:svec-phi} via a topological Wick rotation. 

\begin{rem}
By \cite[Corollary\, 4.39]{KYZZ21}, the boundary-bulk relation holds, i.e.
$$
\FZ_0({}^\svec\svec) \simeq {}^{\FZ_1(\svec)}\svec. 
$$
\end{rem}

\begin{figure}
$$
\begin{tikzpicture}[scale=0.6]
\fill[gray!20] (-3,0) rectangle (3,3) ;
\draw[->-,very thick] (-3,0)--(3,0) node[midway,below] {\footnotesize $\svec$} ;
\draw[->-,very thick] (-3,3)--(-3,0) node[midway,left] {\scriptsize $\svec$} ;
\draw[->-,very thick] (3,3)--(3,0) node[midway,right,scale=0.8] {$\svec$} ;

\draw[dashed] (0,3)--(2.9,0.1) node[near start,left] {\scriptsize $\CY_{m \leftrightarrow e}$}; 

\draw[fill=white] (-3.1,-0.1) rectangle (-2.9,0.1) node[midway,below] {\scriptsize $\svec$} ;
\draw[fill=white] (2.9,-0.1) rectangle (3.1,0.1) node[midway,below] {\scriptsize $\vect$} ;
\node[above] at (-1.2,1) {\scriptsize $\FZ_1(\svec)$} ;
\node[above] at (2,2) {\scriptsize $\FZ_1(\svec)$} ;

\end{tikzpicture}
$$
\caption{This picture depicts a 2d spatial configuration that can realize two 1+1D gapped phases appeared in 1d Kitaev chain and their boundaries via topological Wick rotation. We use $\CY_{m \leftrightarrow e}$ to denote the invertible domain wall associated to the braided auto-equivalence $\FZ_1(\svec) \to \FZ_1(\svec)$ defined by $m\leftrightarrow e$. By \cite{KZ18}, $\CY_{m \leftrightarrow e}$ can be mathematically described by the category $\Fun_{\svec|\svec}(\vect, \vect)$ of $\svec$-$\svec$-bimodule functors. }
\label{fig:svec-phi}
\end{figure}

\subsubsection{The case $\mu=0$ and $t=-\Delta\approx -1$}

This case is obtained from the previous one by making a replacement $\gamma_{j,2}\mapsto\gamma_{j+1,2}$, thus is related to the previous case under the involution $m \leftrightarrow m f$.

In the bulk, there are two topological sectors of states (or bulk excitations) $\one,f$ with the following fusion rules: 
$$\one\otimes\one=f\otimes f=\one, \quad \one\otimes f=f\otimes\one=f.$$
Moreover, we have
$$\hom_{bulk}^{kc2}(\one,\one) = \hom_{bulk}^{kc2}(f,f) = \one \oplus e,$$
$$\hom_{bulk}^{kc2}(\one,f) = \hom_{bulk}^{kc2}(f,\one) = f \oplus m.$$
Comparing them with (\ref{eq:Z1svec-svec-phi}), we see that the observables in this 1d phase form the enriched fusion category ${}_{\quad m \leftrightarrow e}^{\FZ_1(\svec)}\svec$, where the enrichment is twisted from the standard one by the involution $m \leftrightarrow e$. According to Eq.\,(\ref{eq:pic=f_0}) and 
Example\,\ref{expl:nd-SPT} (see also Remark\,\ref{rem:lqs+skeleton}), this 1d phase is the unique non-trivial 1d SPT order with a fermionic $\Zb_2$ onsite symmetry, which is also called a 1d p-wave topological superconductor.

\medskip

When there is a boundary on the left by restricting the model to $i\ge0$, there is only one boundary excitation $\one'$ and we have 
$$\hom_{bdy}^{kc2}(\one',\one') = \one \oplus f.$$
Comparing it with (\ref{eq:svec-vect}), we see that the observables on this boundary form the enriched category ${}^\svec\vect$. The 1d bulk phase and its boundary can be obtained from the 2d spatial configuration depicted in the right half of Figure\,\ref{fig:svec-phi} via a topological Wick rotation. 

\begin{rem}
By \cite[Corollary\, 4.39]{KYZZ21}, the boundary-bulk relation holds, i.e.
$$
\FZ_0({}^\svec\vect) \simeq {}_{\quad m \leftrightarrow e}^{\FZ_1(\svec)}\svec. 
$$
\end{rem}

\begin{rem}
One can construct a domain wall between two 1d gapped phases realized in the Kitaev chain. We leave it as an exercise to show that it can be described mathematically by the enriched category ${}^{\Fun_{\svec|\svec}(\vect, \vect)}\vect$ (see Figure\,\ref{fig:svec-phi}). 
\end{rem}

\begin{rem} \label{rem:Ising=Kitaev}
Note that $\svec=\Rep(\Zb_2)=\vect_{\Zb_2}$ as fusion categories. Therefore, we can simply identify them. As a consequence, we can have the following identifications: 
$$
{}^{\FZ_1(\svec)}\svec = {}^{\FZ_1(\Rep(\Zb_2))}\Rep(\Zb_2) \quad \mbox{and} \quad {}_{\quad m \leftrightarrow e}^{\FZ_1(\svec)}\svec = {}^{\FZ_1(\Rep(\Zb_2))}\vect_{\Zb_2}. 
$$
In other words, enriched fusion categories appeared in Kitaev chain and those appeared in Ising chain are entirely the same. However, the bosonic symmetry charges $\Rep(\Zb_2)$ and the fermionic symmetry charges $\svec$ play different roles in $\FZ_1(\Rep(\Zb_2))$, i.e. 
\begin{align}
\Rep(\Zb_2) &\hookrightarrow \FZ_1(\Rep(\Zb_2)) \quad\quad\quad \quad
\svec \hookrightarrow \FZ_1(\Rep(\Zb_2)) \nn
1, e &\mapsto 1, e \quad\quad\quad\quad\quad\quad\quad \quad\,\,\, 
1,f \mapsto 1,f.   \label{eq:symmetry-charges-embedding}
\end{align}
This difference provides different physical meanings to the topological skeletons in bosonic and fermionic cases. This observation leads us to the classification of all fermionic 1d gapped quantum phases given in Theorem${}^{\mathrm{ph}}$\,\ref{pthm:fermionic-1d-phases}. 
\end{rem}

\section{Classification of 1d gapped quantum phases} \label{sec:outlooks}
In \cite[Section\,7]{KZ21}\cite[Section\ 5.2]{KZ20b}, a unified mathematical framework was proposed for the study of all gapped/gapless quantum liquid phases with/without onsite symmetries in all dimensions. In particular, a quantum liquid $\CX$ can be described by a pair $\CX=(\CX_{\mathrm{lqs}}, \CX_{\mathrm{sk}})$, where $\CX_{\mathrm{lqs}}$ encodes the information of the so-called local quantum symmetry and $\CX_{\mathrm{sk}}$ is the topological skeleton, which is mathematically described by an enriched (higher) category. In a 1+1D chiral (resp. non-chiral) CFT, $\CX_{\mathrm{lqs}}$ is given by a vertex operator algebra (resp. a full field algebra). For gapped phases, $\CX_{\mathrm{lqs}}$ is more subtle. A proper treatment of $\CX_{\mathrm{lqs}}$ requires us to work within the framework of a proper generalization of conformal nets. Indeed, in a recent work \cite{KZ22}, local quantum symmetries and topological skeletons are unified into a single mathematical theory of topological nets. Many subtle issues related to local quantum symmetries are clarified there.

\medskip
In this work, we have focused on the topological skeleton. Although we have checked only two simple lattice models in 1d, we believe that the unifying power of the enriched-category description revealed by these two models is very convincing. It is certainly important to check more known lattice models in higher dimensions. We want to emphasize, however, that it is already interesting and non-trivial to check more 1d models. One can start from more general onsite symmetry given by a finite group $G$. In this case, the categorical symmetry (recall Remark\,\ref{rem:cat-symmetry}) should be given by $\FZ_1(\Rep(G))$ (see also \cite{KZ22}). All possible topological skeletons associated to it can be classified by all fusion categories $\CS$ equipped with a braided equivalence $\phi: \FZ_1(\Rep(G)) \to \FZ_1(\CS)$, or equivalently, by $\CS=(\FZ_1(\Rep(G)))_A$, where $A$ is a Lagrangian algebra in $\FZ_1(\Rep(G))$ and $(\FZ_1(\Rep(G))_A$ denotes the category of right $A$-modules in $\FZ_1(\Rep(G)$. Moreover, the Lagrangian algebras in $\FZ_1(\Rep(G))$ are classified by pairs $(H, \omega)$, where $H$ is a subgroup of $G$ and $\omega\in H^2(H,U(1))$ \cite{Dav10}.  We denote the Lagrangian algebra associated to $(H, \omega)$ by $A_{(H,\omega)}$. As a consequence, we have rediscovered the following well-known classification result. 

\begin{pthm}[\cite{CGW10b,SPGC11}]
All 1d bosonic gapped quantum phases\footnote{All gapped quantum phases in 1d are quantum liquids.} with a finite onsite symmetry are classified by a triple $(G,H,\omega)$, where $G$ is the onsite symmetry defined by a finite group, $H$ is a subgroup of $G$ and $\omega \in H^2(H,U(1))$ is a 2-cocycle. 
\end{pthm}

Moreover, the general theory in \cite{KZ20b} provides us with the following new result begging to be checked in concrete lattice models. 
\begin{pthm} \label{pthm:bosonic-1d-phases}
The topological skeleton of the 1d bosonic gapped phase associated to $(G,H,\omega)$ is given by the enriched fusion category ${}^{\FZ_1(\Rep(G))}((\FZ_1(\Rep(G)))_{A_{(H,\omega)}})$. 
\end{pthm}

\begin{rem}
In the well-known classification \cite{CGW10b,SPGC11}, $G$ is the symmetry group of the Hamiltonian, and the ground state breaks the symmetry $G$ to a subgroup $H$. This coincides precisely with the fact that the vacuum sector of states provides a condensation of the Lagrangian algebra $A_{(H,\omega)}$ in the categorical symmetry $\FZ_1(\Rep(G))$ (recall Remark\,\ref{rem:condensation-1} and \ref{rem:condensation-2}). 
\end{rem}

We use the pair $(G,z)$ to denote a fermionic finite onsite symmetry. In particular, $G$ is a finite group and $z$ is an element in the center of $G$ defining the fermion parity. We denote the category of $G$-representations  equipped with a new braiding structure that are compatible with the fermion parity by $\Rep(G,z)$. Since $\Rep(G)=\Rep(G,z)$ as fusion categories, we can also identify their Drinfeld centers, i.e. $\FZ_1(\Rep(G))=\FZ_1(\Rep(G,z))$. But keep in mind that bosonic symmetry charges $\Rep(G)$ and fermionic symmetry charges $\Rep(G,z)$ are embedded in $\FZ_1(\Rep(G))$ differently (recall the $\Zb_2$-case (\ref{eq:symmetry-charges-embedding})). We immediately recover the classification of all 1d gapped quantum phases with onsite fermionic symmetry $(G,z)$, a result which was first obtained in \cite{CGW11b} based on the idea that fermionic systems and spin systems can be mapped to each other via the Jordan-Wigner transformations. 

\begin{pthm} \label{pthm:fermionic-1d-phases}
All 1d gapped quantum phases with a finite fermionic onsite symmetry $(G,z)$ are classified by the same triples $(G,H,\omega)$ as the bosonic cases, and the associated topological skeleton is ${}^{\FZ_1(\Rep(G))}((\FZ_1(\Rep(G)))_{A_{(H,\omega)}})$, where  $A_{(H,\omega)}$ is a Lagrangian algebra in $\FZ_1(\Rep(G))$. The fermionic symmetry charges are embedded in $\FZ_1(\Rep(G))$ canonically according to $\Rep(G,z) \hookrightarrow \FZ_1(\Rep(G,z))=\FZ_1(\Rep(G))$. 
\end{pthm}

\begin{rem}
Each topological skeleton ${}^{\FZ_1(\Rep(G))}((\FZ_1(\Rep(G)))_{A_{(H,\omega)}})$ is associated to two 1d gapped quantum phases. One is bosonic, and the other is fermionic. Note that, in the fermionic cases, the fermion parity never breaks by a boson condensation. 
\end{rem}

It was conjectured in \cite{JW20,KLWZZ20b} that there should exist two lattice models realizing in low energy two given and Morita equivalent algebraic higher symmetries (recall Remark\,\ref{rem:cat-symmetry}), respectively, such that they are actually dual to each other, i.e. existing an explicit non-local duality transformation mapping one model to the other and symmetric local operators to symmetric local operators. By Theorem${}^{\mathrm{ph}}$\,\ref{pthm:bosonic-1d-phases} and \ref{pthm:bosonic-1d-phases}, this conjecture says, in particular, that there are lattice models realizing all 1d gapped phases with a fixed finite onsite symmetry $G$, respectively, such that they are dual to each other. When $G=\Zb_2$, two from the Ising chain (i.e. the $B=0$ case and the $J=0$ case) and two from the Kitaev chain are actually dual to each other via the Kramers-Wannier duality transformation and the Jordan-Wigner transformation. It is an interesting project to provide a systematic construction of appropriate 1d lattice models and the duality transformations for an arbitrary $G$ (see an interesting and related work \cite{LDOV21}).

\begin{rem} \label{rem:terminology} 
The appearance of the enriched category ${}^\CB\CS$ is rare in literature, but $\CB$ and $\CS$ have appeared in various contexts under different names. The category $\CS$ of TDL's was called an ``algebraic higher symmetry'' in \cite{JW20,KLWZZ20b}, but was called a ``fusion categorical symmetry'' in \cite{TW19} and was called a ``categorical symmetry'' by many others. Before we apply the topological Wick rotation, the category $\CB$ can be viewed as the bulk of the gapped boundary phase $\CS$, and was called the ``categorical symmetry'' in \cite{JW20,KLWZZ20b}. 
\end{rem}

\begin{rem}
The importance of enriched fusion categories in the study of topological phases was discovered in the study of gapless phases \cite{KZ20a,KZ21}. Its higher dimensional analogues were proposed to give a unified framework to study all gapped/gapless liquid phases with/without symmetries \cite{KZ20a,KZ21,KZ20b}. Its relevance in the study of topological phase transitions was demonstrated in \cite{CJKYZ20}. However, its significance has not yet been recognized by condensed matter theorists. Perhaps a partial reason for this delay is the abstractness of the categorical language. We hope that through the study of two simple and well-known lattice models in this work we can help some readers to break the language barrier. 
\end{rem}

\appendix
\section{Enriched categories} \label{sec:enriched-categories}
We briefly explain all mathematical results that are needed in this work. 

\medskip
Given a fusion category $\CC$ and a finite simple left $\CC$-module $\CM$ with a left $\CC$-action $\odot: \CC \times \CM \to \CM$, there is a $\CC$-enriched category  ${}^\CC\CM$ obtained from the so-called canonical construction \cite{Kel67}. The objects in ${}^\CC\CM$ are precisely objects in $\CM$. The hom spaces $\hom_{{}^\CC\CM}(x,y)$ for $x,y\in\CM$ are given by the so-called internal hom: $[x,y]$, which is defined by the following conditions:
$$
\hom_\CM(a\odot x, y) \simeq \hom_\CC(a,[x,y]), \quad\quad \forall a\in\CC,x,y\in\CM. 
$$
The hom spaces $\hom_{{}^\CC\CM}(x,y)=[x,y] \in \CC$ for $x,y\in\CM$ determines the structure of ${}^\CC\CM$ completely. In particular, the composition of morphisms and identity morphisms: 
\be \label{eq:comp-morphism}
[y,z] \otimes [x,y] \rightarrow [x,z] \quad \mbox{and} \quad \one_\CC \to [x,x]
\ee
are naturally induced by the universal property of the internal homs. If, in addition, $\CC$ is braided, and $\CM$ is monoidal and is equipped with a braided functor $\phi: \CC \to \FZ_1(\CM)$, then ${}^\CC\CM$ becomes a $\CC$-enriched fusion 1-category \cite{MP19,KZ20a}.

\medskip
Let $\hilb$ be the category of finite dimensional vector spaces. Let $\Rep(\Zb_2)$ be the category of finite dimensional $\Zb_2$-representations and $\hilb_{\Zb_2}$ the category of $\Zb_2$-graded finite dimensional vector spaces. Let $\svec$ be the category of finite dimensional super vector spaces. Note that $\Rep(\Zb_2)$, $\hilb_{\Zb_2}$ and $\svec$ are all equivalent as fusion categories. Therefore, their Drinfeld centers can be identified, i.e. 
$$
\FZ_1(\Rep(\Zb_2))=\FZ_1(\hilb_{\Zb_2}) = \FZ_1(\svec). 
$$
We denote the only simple object of $\hilb$ by $\one$. We denote the two simple objects of $\Rep(\Zb_2)$ by $\one, e$, and those of $\hilb_{\Zb_2}$ by $\one, m$, and those of $\svec$ by $\one, f$, and those of $\FZ_1(\Rep(\Zb_2))$ by $\one, e, m, f$ (i.e. the same four simple anyons in 2d toric code model). These notations are justified by three different braided embeddings: 
\begin{align} \label{eq:three-embedding}
&\Rep(\Zb_2) \hookrightarrow \FZ_1(\Rep(\Zb_2)) \quad\quad 
\hilb_{\Zb_2} \hookrightarrow \FZ_1(\Rep(\Zb_2)) \quad\quad  \svec \hookrightarrow \FZ_1(\Rep(\Zb_2)) \\
&\one \mapsto \one, e \mapsto e, \quad\quad\quad\quad\quad\quad\quad 
\one \mapsto \one, m \mapsto m, \quad\quad\quad\quad\quad
\one \mapsto \one, f\mapsto f. 
\end{align} 
The non-trivial fusion rules of $\FZ_1(\Rep(\Zb_2))$ are $e\otimes e=m\otimes m=f\otimes f=\one$ and $e\otimes m=f$. 

\medskip
Now we give a few examples of enriched (fusion) categories all obtained from canonical constructions. All of them are used in this work. 
\bnu
\item ${}^{\FZ_1(\Rep(\Zb_2))}\Rep(\Zb_2)$: 
\begin{align} \label{eq:Z(RepZ2)-RepZ2}
&[\one,\one]=\one\oplus m, \quad\quad [\one, e] = [e,\one] = e\oplus f, \quad\quad [e,e]=\one\oplus m. \end{align}

\item ${}^{\FZ_1(\Rep(\Zb_2))}\hilb_{\Zb_2}$: 
\be \label{eq:Z(RepZ2)-HilbZ2}
[\one,\one]=\one\oplus e, \quad\quad [\one,m]=[m,\one]= m\oplus f, \quad\quad [m,m]=\one\oplus e. 
\ee

\item ${}^{\Rep(\Zb_2)}\Rep(\Zb_2)$: 
\be \label{eq:RepZ2-RepZ2}
[\one,\one]=\one, \quad\quad [\one,e]=[e,\one]=e, \quad\quad [e,e]=\one. 
\ee

\item ${}^{\hilb_{\Zb_2}}\hilb_{\Zb_2}$: 
\be \label{eq:HilbZ2-HilbZ2}
[\one,\one]=\one, \quad\quad [\one,m]=[m,\one]=m, \quad\quad [m,m]=\one.
\ee

\item ${}^{\Rep(\Zb_2)}\hilb$: 
\be \label{eq:RepZ2-Hilb}
[\one, \one] = \one \oplus e. 
\ee

\item ${}^{\hilb_{\Zb_2}}\hilb$: 
\be \label{eq:HilbZ2-Hilb}
[\one, \one] = \one \oplus m. 
\ee

\item ${}^{\FZ_1(\svec)}\svec$: 
\be \label{eq:Z1svec-svec}
[\one,\one] = [f,f] =\one \oplus m, \quad\quad [\one,f] = [f,\one] = f\oplus e. 
\ee

\item ${}_{\quad m \leftrightarrow e}^{\FZ_1(\svec)}\svec$ with the enrichment twisted by the non-trivial braided auto-equivalence of $\FZ_1(\svec)$ defined by $m \leftrightarrow e$. 
\be \label{eq:Z1svec-svec-phi}
[\one,\one] = [f,f] =\one \oplus e, \quad\quad [\one,f] = [f,\one] = f\oplus m. 
\ee

\item ${}^{\svec}\svec$: 
\be \label{eq:svec-svec}
[\one,\one] = [f,f] =\one, \quad\quad [\one,f]=[f,\one]=f. 
\ee

\item ${}^\svec\vect$: 
\be \label{eq:svec-vect}
[\one,\one] = \one \oplus f. 
\ee

\item ${}^{\FZ_1(\Rep(\Zb_2))}\hilb$: 
\be \label{eq:ZZ2-vect}
[\one,\one] = \one \oplus m \oplus e \oplus f. 
\ee
\enu

\begin{rem}
This paper is written for physicists. In order to keep the paper not too mathematically technical, we decide to hide some technical parts in Remarks. In Eq.\,(\ref{eq:Z(RepZ2)-RepZ2})-(\ref{eq:ZZ2-vect}), we have only presented all internal homs as objects (recall (\ref{eq:comp-morphism}). We have not given the the identity morphisms and the compositions of morphisms. In this remark, we illustrate them in a single example: the enriched category ${}^{\FZ_1(\Rep(\Zb_2))}\Rep(\Zb_2)$: 
\bnu
\item Identity morphisms: 
\be \label{eq:identity}
\one \xrightarrow{1_\one \oplus \, 0} \one\oplus m = [\one,\one] = [e,e].
\ee
Since the only morphism from $\one$ to $m$ is the zero morphism, we will not spell out this type of zero morphisms explicitly from now on.

\item Compositions of morphisms: 
\begin{align}
&[\one,\one] \otimes [\one,\one] = (\one\oplus \one) \oplus (m \oplus m) \xrightarrow{ (1_\one\oplus 1_\one) \oplus (1_m \oplus 1_m)} \one \oplus m = [\one,\one],  \label{eq:comp-1} \\
&[\one,e]\otimes[e,\one] = (\one\oplus \one) \oplus (m \oplus m) \xrightarrow{ (1_\one\oplus 1_\one) \oplus (1_m \oplus 1_m)} \one \oplus m =[e,e], \label{eq:comp-2} \\
&[e,\one]\otimes[\one,e] = (\one\oplus \one) \oplus (m \oplus m) \xrightarrow{ (1_\one\oplus 1_\one) \oplus (1_m \oplus 1_m)} \one \oplus m =[\one,\one],  \label{eq:comp-3} \\
& [e,e] \otimes [e,e] = (\one\oplus \one) \oplus (m \oplus m) \xrightarrow{ (1_\one\oplus 1_\one) \oplus (1_m \oplus 1_m)} \one \oplus m = [e,e].  \label{eq:comp-4}
\end{align}
\enu
Moreover, since the enriched category ${}^{\FZ_1(\Rep(\Zb_2))}\Rep(\Zb_2)$ is also monoidal, it has another defining data: the horizontal fusion morphism $[x',y'] \otimes [x,y] \to [x'\otimes x, y'\otimes y]$, which is 
canonically induced from the following morphism (via the universal property of the internal hom $[x'\otimes x, y'\otimes y]$):
$$
([x',y']\otimes [x,y]) \odot (x'\otimes x) \xrightarrow{\simeq} ([x',y'] \odot x') \otimes ([x,y] \odot x) \rightarrow y' \otimes y, 
$$
where ``$\simeq$'' uses a half-braiding to exchange $[x,y]$ with $x'$ and the second morphism is defined by the universal morphisms of the internal homs $[x',y']$ and $[x,y]$. Explicit computation gives the following horizontal fusion morphisms: 
\begin{align} \label{eq:horizontal-fusion}
&[\one,\one] \otimes [\one,\one] = (\one\oplus \one) \oplus (m \oplus m) \xrightarrow{(1_\one\oplus 1_\one) \oplus (1_m \oplus 1_m)} 
\one \oplus m = [\one,\one]. 
\end{align}
We leave the rest of horizontal fusion morphisms as exercises. 
\end{rem}

\end{document}